\documentclass[12pt,epsf]{article}
\usepackage{graphicx}

\usepackage{color}

\begin{document}

%%%%%%%%%%%%%%%%%%%%%%%%%%%%%%%%%%%%%%%%%%%%%%%%%%%%%%%%%%%%%%%%%%%%%%%
\newcommand{\gtrsim}{ \mathop{}_{\textstyle \sim}^{\textstyle >} }
\newcommand{\lesssim}{ \mathop{}_{\textstyle \sim}^{\textstyle <} }

%%%%%%%%%%%%%%%%%%%%%%%%%%%%%%%%%%%%%%%%%%%%%%%%%%%%%%%%%%%%%%%%%%%%%%%

\baselineskip 0.7cm

\begin{titlepage}

\begin{flushright}
ICRR-Report-643-2012-32\\
IPMU13-0009\\
UT-13-01\\
\end{flushright}

\vskip 2.35cm

\begin{center}
{\large \bf 
  Axions : Theory and Cosmological Role 
}
\vskip 1.2cm
Masahiro Kawasaki$^{(a,b)}$
and
Kazunori Nakayama$^{(c,b)}$

\vskip 0.4cm

{\it $^a$Institute for Cosmic Ray Research, University of Tokyo, Kashiwa 277-8582, Japan}\\
{\it $^b$Institute for the Physics and Mathematics of the universe,
University of Tokyo, Kashiwa 277-8568, Japan}\\
{\it $^c$Department of Physics, University of Tokyo, Tokyo 113-0033, Japan}\\

\vskip 1.5cm

\abstract{ 

We review recent developments on axion cosmology.
Topics include : axion cold dark matter, axions from topological defects, axion isocurvature perturbation and
its non-Gaussianity and axino/saxion cosmology in supersymmetric axion model.
 
}

\end{center}
\end{titlepage}

\setcounter{page}{2}

\tableofcontents

%%%%%%%%%%%%%%%%%%%%%%%%%%%%%%%%%%%%%%%%%%%%%%%%%%%%%%%%%%%%%%%%%%%%%%%%%%%

%%%%%%%%%%%%%%%%%%%%%%%%%%%%%%%%%%%%%
\section{Introduction}
%%%%%%%%%%%%%%%%%%%%%%%%%%%%%%%%%%%%%

The standard model (SM) in particle physics is well established after the discovery of the
Higgs-boson like particle at LHC~\cite{:2012gk,:2012gu}.
While it goes without saying that SM is a successful theory, it suffers from the strong CP problem~\cite{Kim:1986ax,Kim:2008hd}.
The following term in the Lagrangian allowed by the gauge symmetry,
\begin{equation}
	\mathcal L = \theta \frac{g_s^2}{32\pi^2}G_{\mu\nu}^a \tilde G^{\mu\nu a},  \label{theta}
\end{equation}
violates CP and contributes to the neutron electric dipole moment (NEDM).
Here $g_s$ is the QCD gauge coupling constant, $G_{\mu\nu}^a$ is the gluon field strength, $\tilde G_{\mu\nu}^a$ is its dual,
and $\theta$ is a constant parameter.
Recent experimental bound on the NEDM reads $|d_n| < 2.9\times 10^{-26}e$\,cm (90\% CL)~\cite{Baker:2006ts}.
This leads to the constraint on the $\theta$ parameter as $\theta < 0.7\times 10^{-11}$~\cite{Kim:1986ax,Kim:2008hd}.
There seems to be no reason in the SM why $\theta$ must be so small : this is the strong CP problem.

Peccei and Quinn~\cite{Peccei:1977hh,Peccei:1977ur} proposed a beautiful solution to the strong CP problem.
They introduced anomalous global U(1) symmetry, which we denote by U(1)$_{\rm PQ}$ called PQ symmetry,
which is spontaneously broken.
Then the $\theta$ term is replaced by a dynamical field, which automatically goes to zero by minimizing the potential.
It was soon realized that such a solution to the strong CP problem leads to a
light pseudo scalar particle : axion~\cite{Weinberg:1977ma,Wilczek:1977pj}.
The axion is a pseudo-Nambu Goldstone boson in association with spontaneous breakdown of the PQ symmetry.
It has a coupling as 
\begin{equation}
	\mathcal L = \frac{g_s^2}{32\pi^2}\frac{a}{F_a} G_{\mu\nu}^a \tilde G^{\mu\nu a},  \label{aGG}
\end{equation}
where $a$ denotes the axion field and $F_a$ is the scale of PQ symmetry breaking.
Then the $\theta$ parameter is effectively replaced with $\theta + a/F_a$.
It has a CP-conserving potential minimum at $\theta + a/F_a = 0$, hence the CP angle is dynamically tuned to be zero
without fine-tuning.
It is a very attractive idea for solving the strong CP problem.

Once we believe the PQ solution to the strong CP problem, the axion may play an important role
in particle phenomenology and cosmology.
In this article we review the axion cosmology, in particular focusing on recent developments in the last few years.
We do not aim to explain underlying physics of the axion, for which
we refer to excellent reviews~\cite{Kim:1986ax,Kim:2008hd}.
%Although there are already excellent reviews on the physics of axions~\cite{Kim:1986ax,Kim:2008hd},
%we aim to focus on recent developments with an emphasis on cosmological issues on the axion in this article.
In Section~\ref{sec:PQ}, the PQ and axion models, and experimental/observational constraints are briefly summarized.
In Section~\ref{sec:cosmology}, we discuss the axion cosmology.
In particular, recent calculations on the cold and hot axion abundances, axions emitted from topological defects,
and axion isocurvature fluctuation and its non-Gaussianity are summarized.
In Section~\ref{sec:SUSY}, we focus on supersymmetric (SUSY) axion models and their cosmological effects.
There are some recent developments on the evaluation of the saxion and axino abundances.
In Section~\ref{sec:conc}, we mention some related topics which are not covered in the main text.

%%%%%%%%%%%%%%%%%%%%%%%%%%%%%%%%%%%%%
\section{Peccei-Quinn mechanism and axion}  \label{sec:PQ}
%%%%%%%%%%%%%%%%%%%%%%%%%%%%%%%%%%%%%

%%%%%%%%%%%%%%%%%%%%%%%%%%%%%%%%%%%%%
\subsection{Models of invisible axion}
%%%%%%%%%%%%%%%%%%%%%%%%%%%%%%%%%%%%%

Early models of axion~\cite{Weinberg:1977ma,Wilczek:1977pj}, where the axion was
associated with the weak scale Higgs boson, were soon ruled out experimentally.
Currently the most axion models 
make the axion invisible by assuming very high PQ scale.
There are two known class of invisible axion models :
KSVZ model (or also called hadronic axion model)~\cite{Kim:1979if,Shifman:1979if}
and DFSZ model~\cite{Dine:1981rt,Zhitnitsky:1980tq}.

In the KSVZ model~\cite{Kim:1979if,Shifman:1979if}, heavy quark pair, $Q$ and $\bar Q$ are introduced which 
are fundamental and anti-fundamental representations of SU(3)$_c$ and couple to the PQ scalar $\phi$ as
\begin{equation}
	\mathcal L = k\phi Q\bar Q.
\end{equation}
Here U(1)$_{\rm PQ}$ charges are assigned as $\phi(+1)$, $Q (-1/2)$ and $\bar Q (-1/2)$.
These quarks become heavy after $\phi$ gets a vacuum expectation value (VEV) of $\eta$,
%\footnote{
which is related to the PQ scale $F_a$ defined in (\ref{aGG}),
through the relation $F_a = \eta / N_{\rm DW}$. 
Here a model-dependent integer $N_{\rm DW}$ is called the domain wall number 
(See Section~\ref{sec:cosmology}).
%}
All the SM fields are assumed to be singlets under U(1)$_{\rm PQ}$.
Clearly, this global U(1)$_{\rm PQ}$ has an anomaly under the QCD.
Therefore, the axion obtains a coupling as Equation \ref{aGG} 
and the theta angle is dynamically tuned to be zero by the PQ mechanism.
For a minimal case where only one pair of heavy quarks is introduced, we have $N_{\rm DW}=1$. 
There is no domain wall problem in this case. (See Section~\ref{sec:cosmology}).

In the DFSZ model~\cite{Dine:1981rt,Zhitnitsky:1980tq}, the PQ field couples to the SM Higgs.
In this model two Higgs doublets are required : $H_1$ and $H_2$.
We assume that $H_1$ transforms as the SM Higgs and $H_2$ as its conjugation under the SM gauge groups.
Moreover, PQ charges are assigned so that the combination $H_1H_2$ has $-1$.
%(For simplicity, we assume $H_1 (H_2)$ has PQ charge $-1 (0)$ hereafter.)
Then we can write down the interaction term among them in the potential of PQ and Higgs sector,
\begin{equation}
	-\mathcal L = |\phi|^2(c_1 |H_1|^2+c_2|H_2|^2) + c_3 |H_1|^2|H_2|^2 + c_4 |H_1 H_2|^2 + (\mu \phi H_1 H_2 + {\rm h.c.}),
	%+\kappa_1\left(|H_1|^2 -v_1\right)^2+\kappa_2\left(|H_2|^2 -v_2\right)^2 + \kappa \left(|\phi|^2- F_a^2 \right)^2
	\label{LDFSZ}
\end{equation}
where $c_1,\dots, c_4$ are numerical constants and $\mu$ is a dimensional parameter.
Let us suppose that the VEVs of $H_1$ and $H_2$ give up- and down-type quark masses.
Then up-type SM quarks necessarily have PQ charge and the U(1)$_{\rm PQ}$ becomes anomalous under the QCD.
Hence the axion coupling like Equation \ref{aGG} appears, solving the strong CP problem.
In this case, the domain wall number is calculated as $N_{\rm DW} = 3$ reflecting three family of SM quarks.\footnote{
	This depends on PQ charge assignments on the Higgs field.
	For example, if we take the PQ charge of $H_1H_2$ to be $-2$, the allowed term is $\phi^2 H_1H_2$
	and we obtain $N_{\rm DW}=6$.
}
Hence it may suffer from the domain wall problem if the PQ symmetry is broken after inflation.
A distinct feature of the DFSZ model is that SM fermions, including leptons, have tree-level coupling to the axion
through the Higgs-axion mixing.

In the so-called variant axion model~\cite{Peccei:1986pn,Krauss:1986wx}, 
it is assumed that $H_1$ only couples to one family of up-type quarks (say, top quark).
All other quarks have zero PQ charges and obtain masses from $H_2$.
In this case, we have $N_{\rm DW}=1$ and there is no domain wall problem.
See \cite{Chen:2010su} for implications of Higgs sector in this model at collider experiments.

%%%%%%%%%%%%%%%%%%%%%%%%%%%%%%%%%%%%%
\subsection{Astrophysical and experimental constraints}
\label{sec:astro}
%%%%%%%%%%%%%%%%%%%%%%%%%%%%%%%%%%%%%

Since the axion interaction is very weak and much lighter than typical temperature of stars,
axions are emitted from stars.
Hence the PQ scale is bounded below so as not to change evolutions of stars significantly~\cite{Raffelt:1996wa,Raffelt:1999tx}.
The most stringent constraint comes from the observation of SN1987A.
In order for the axion emission not to shorten the burst duration, the PQ scale is bounded as $F_a \gtrsim 4\times 10^8$\,GeV.
This bound relies on the axion-hadron interaction, which always exists in the axion model solving the strong CP problem.\footnote{
	For much smaller $F_a$ the axion becomes optically thick and the axion emission rate is suppressed.
	The bound from the burst duration of SN1987A disappears at $F_a \lesssim 10^6$\,GeV.
	On the other hand, for even smaller $F_a$, the detection rate of the thermally emitted axion from SN1987A at the Kamiokande
	would become large. As a result, the SN1987A does not pose a constraint for $10^5$\,GeV $\lesssim F_a \lesssim 10^6$\,GeV.
}
The observation of horizontal branch (HB) stars in globular clusters also set lower bound on the PQ scale as $F_a \gtrsim 10^7$\,GeV.\footnote{
	The bound from HB stars come from axion-photon interaction, which is model-dependent.
	If the coefficient of the axion-photon-photon interaction is somehow chosen to be much smaller than unity, 
	the bound from HB star cooling can be avoided. This, combined with the argument from SN1987A, 
	leads to the so-called ``hadronic axion window''
	at $F_a \sim 10^6$\,GeV ($m_a \sim 1$\,eV) where constraints from stars are absent~\cite{Chang:1993gm,Moroi:1998qs}.
	(In the DFSZ model, the axion-electron interaction induces rapid star cooling for this value of $F_a$, hence there is no
	such an window.)
	However, recent cosmological data disfavor such a scenario, because thermally produced axions 
	contribute as hot dark matter component of the Universe. (See Section~\ref{sec:hot_axion}). 
}

There are activities on experimental searches for the axion.
\paragraph{Axion helioscope}
Axion helioscopes try to detect axions with energy of order keV emitted from the center of the Sun 
through the axion-photon conversion process under the magnetic field~\cite{Sikivie:1983ip}.
Currently the CAST experiment~\cite{Arik:2008mq,Arik:2011rx} put the most stringent bound on the 
strength of axion-photon-photon coupling in this way.
In particular, the resonant conversion expected from the plasma frequency of photon induced by the buffer gas
can improve the sensitivity and they begin to cover the parameter region predicted in the QCD axion for $F_a \sim 10^{7-8}$\,GeV.
The Tokyo axion helioscope also reaches the QCD axion prediction for small range of the axion mass~\cite{Inoue:2008zp}.
Future project, called IAXO~\cite{Irastorza:2011gs}, may reveal the QCD axion for $F_a\sim 10^9$\,GeV.

\paragraph{Axion haloscope}
Axion haloscopes try to detect dark matter (DM) axions in the Galaxy by using the microwave cavity~\cite{Sikivie:1983ip,Bradley:2003kg}.
Under the magnetic field, the axion DM may produce radio wave with its frequency corresponding to the axion mass,
and it is amplified if the size of the cavity matches with the Compton wave length of the axion.
Note that this technique crucially relies on the assumption that the observed DM consists of cold axion. 
The ADMX experiment already begins to exclude axion DM for a limited range of the axion mass
around $F_a\sim 10^{11}$\,GeV~\cite{Asztalos:2009yp}.
It is expected to cover the wide range of parameters consistent with axion DM.

\paragraph{Laser searches}
Axion mixes with the photon in the external magnetic field due to the axion-photon-photon interaction term.
The ``light shining through a wall''  experiments utilize the laser, 
a part of whose light would pass through a wall as axions
under the magnetic field. The ALPS collaboration placed the most strict bound by this idea~\cite{Ehret:2010mh},
although currently it does not reach the CAST sensitivity and the bound from HB star cooling.
Ideas to significantly improve the sensitivity of laser search is proposed~\cite{Sikivie:2007qm}.

\paragraph{Long range forces}
Since axion is very light, it can mediate macroscopic long range forces~\cite{Moody:1984ba}.
Due to the CP-odd nature of the axion, the coherent interaction (or monopole-monopole interaction) 
between macroscopic objects are suppressed. 
Instead, searches for monopole-dipole interactions provide limits on
the axion-nucleon and axion-electron coupling~\cite{Heckel:2006ww,Hoedl:2011zz},
although it is still far from the prediction of QCD axion.\footnote{
	It is argued that searches for monopole-monopole force combined with star cooling constraints
	give stronger constraint~\cite{Raffelt:2012sp}.
}

To summarize, the best limit on the PQ scale and the axion mass comes from astrophysical arguments.
Future axion helioscope experiment may reach a realistic parameters predicted in the QCD axion model.
The cavity experiment will also be sensitive to the axion DM : if axion is the dominant component of DM,
the ADMX will detect its signatures for realistic values of $F_a$.
Some novel ideas are also proposed for detecting the axion DM for $F_a \gtrsim 10^{15}$\,GeV~\cite{Graham:2011qk}.

%%%%%%%%%%%%%%%%%%%%%%%%%%%%%%%%%%%%%
\section{Axion cosmology}
\label{sec:cosmology}
%%%%%%%%%%%%%%%%%%%%%%%%%%%%%%%%%%%%%

%%%%%%%%%%%%%%%%%%%%%%%%%%%%%%%%%%%%%
\subsection{Evolution of PQ scalar}
\label{sec:PQ-filed_evolution}
%%%%%%%%%%%%%%%%%%%%%%%%%%%%%%%%%%%%%

\begin{figure}
\centering
\includegraphics [width = 10cm, clip]{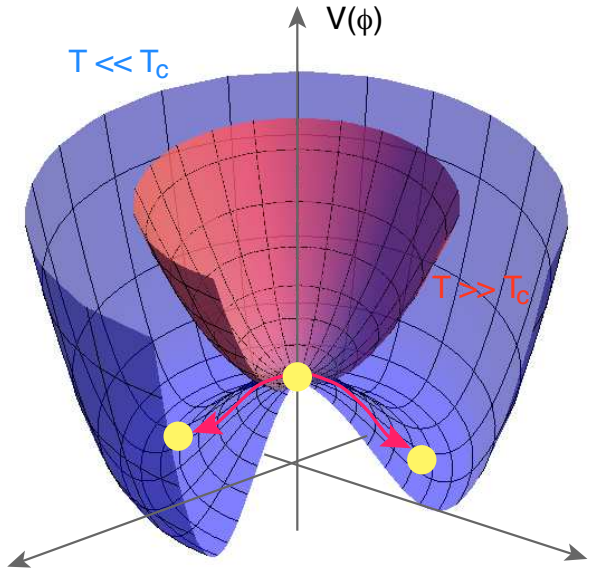}
\caption{Potential of the PQ scalar.}
\label{figure1}
\end{figure}

The Peccei-Quinn (PQ) scalar field has the following lagrangian:
\begin{equation}
   {\cal L} = \frac{1}{2} |\partial_\mu \phi|^2 - V_{\rm eff}(\phi,T),
\end{equation}
where $V(\phi,T)$ is the effective potential at temperature $T$ and given by
\begin{equation}
   V_{\rm eff} = \frac{\lambda}{4} (|\phi|^2 - \eta^2)^2
   + \frac{\lambda}{6}T^2|\phi|^2.
   \label{eq:PQpot}
\end{equation}
The above lagrangian is invariant under the global $U(1)_{\rm PQ}$ transformation,
$\phi \rightarrow \phi e^{i\alpha}$ with $\alpha$ constant.
At high temperature $T> T_c \equiv \sqrt{3}\eta$, the potential has the minimum 
at $\phi =0$ and the vacuum has the $U(1)_{\rm PQ}$ symmetry (Figure~\ref{figure1}).
However, as the cosmic temperature decreases, the vacuum with $\phi=0$ becomes 
unstable and the PQ scalar $\phi$ obtains vacuum expectation value 
$|\phi| =\eta$ (Figure~\ref{figure1}).
Thus, $U(1)_{\rm PQ}$ symmetry is spontaneously brokem at $ T< T_c$.
The axion $a$ is a Nambu-Goldstone boson associated with this spontaneous symmetry 
breaking and corresponds to the phase direction of the PQ scalar as
\begin{equation}
   \phi = |\phi|e^{i\theta_a} = |\phi|e^{ia/\eta}.
\end{equation}
Since the phase direction is flat in the potential~\ref{eq:PQpot}, 
the axion is massless at this point. 

%%%%%%%%%%%%%%%%%%%%%%% FIGURE  %%%%%%%%%%%%%%%%%%%%%%%%%%%%%%%%%%%%%%
%%%%%%%%%%%%%%%%%%%%%%%%%%%%%%%%%%%%%%%%%%%%%%%%%%%%%%%%%%%%%%%%%%%%%%

When the $U(1)_{\rm PQ}$ symmetry is spontaneously broken, one-dimensional 
topological defects called axionic strings are formed. 
After formation the axionic string networks evolve by emitting axions 
and follow the scaling solution.
The emission of axions from the strings and its contribution to the present density 
is described in Section~\ref{sec:axionic_string}. 

When the cosmic temperature further decreases and becomes comparable 
to the QCD scale $\Lambda$ 
($\simeq 0.1$~GeV), the axion obtains its mass through the QCD non-perturbative 
effect. 
Then, the axion potential is written as
\begin{equation}
   V(a) = \frac{m_a^2\eta^2}{N_{\rm DW}^2}
   \left( 1 - \cos\frac{N_{\rm DW} a}{\eta}\right),
   \label{eq:axion-pot}
\end{equation}
where $m_a$ is the mass of the axion.
Here $N_{\rm DW}$ is called the domain wall number which is a model-dependent integer 
related to the color anomaly.
The axion mass $m_a$ depends on the temperature and it was first calculated 
in References~\cite{Gross:1980br}. 
More recently, Wantz and Shellard~\cite{Wantz:2009it} obtained the temperature 
dependence of $m_a$ using the interacting instanton liquid model~\cite{Wantz:2009mi}. 
They gave a simple approximation in the form of the power-law as
\begin{equation}
   m_a(T) = \left\{
   \begin{array}{ll}
      4.05\times 10^{-4}\frac{\Lambda^2}{F_a} 
      \left(\frac{T}{\Lambda}\right)^{-3.34} & T > 0.26\Lambda ,\\
      3.82\times 10^{-2}\frac{\Lambda^2}{F_a} & T < 0.26\Lambda,
   \end{array}\right.
   \label{eq:axion-mass}
\end{equation}
where $\Lambda\simeq 400$MeV and $F_a$ is the axion decay constant given by 
$F_a = \eta/N_{\rm DW}$.
The axion potential~\ref{eq:axion-pot} explicitly breaks $U(1)_{\rm PQ}$ symmetry 
into its subgroup $Z_{N_{\rm DW}}$ and has degenerated minima (vacua)  
at $a = 0, 2\pi/N_{\rm DW},\cdots 2\pi(N_{\rm DW}-1)/N_{\rm DW}$.
When the Hubble parameter $H$ becomes comparable to the axion mass $m_a$, 
the axion  starts to roll down to one of the minima. 
Since the axion field settles into different minima in different places of the 
Universe, domain walls are formed between different vacua.  
These domain walls attach the axionic strings.
For $N_{\rm DW}=1$ the domain walls are disk-like objects whose boundaries 
are axionic strings and collapse by their tension.
On the other hand, the string-domain wall networks are very complicated for
$N_{\rm DW}\ge 2$ and dominate the Universe soon after their formation,
which causes a serious cosmological difficulty called domain wall problem.  
More details about the axion domain walls are found in 
Sections~\ref{sec:axionic_wall} and \ref{sec:axionic_wall_N}.

%%%%%%%%%%%%%%%%%%%%%%%%%%%%%%%%%%%%%
\subsection{Abundance of the axion}
%%%%%%%%%%%%%%%%%%%%%%%%%%%%%%%%%%%%%

%%%%%%%%%%%%%%%%%%%%%%%%%%%%%%%%%%%%%
\subsubsection{Cold axion}
%%%%%%%%%%%%%%%%%%%%%%%%%%%%%%%%%%%%%

The axion field starts to oscillate when the cosmic expansion rate $H$ becomes
comparable to the axion mass $m_a$. 
Using Equation~\ref{eq:axion-mass}, the condition $m_a(T_1)= 3H(T_1)$ leads to
\begin{equation}
  T_1 = 0.98~{\rm GeV} \left(\frac{F_a}{10^{12}{\rm GeV}}\right)^{-0.19}
  \left(\frac{\Lambda}{400{\rm MeV}}\right).
\end{equation}
The oscillation of the coherent field $a$ is described by the equation of motion,
\begin{equation}
  \ddot{a} + 3\frac{\dot{R}}{R} \dot{a} + m^2(T) a = 0,
  \label{eq:eq_of_motion}
\end{equation}
where $R$ is the scale factor. 
From Equation~\ref{eq:eq_of_motion} we get
\begin{equation}
   \dot{\rho}_a = -3\frac{\dot{R}}{R}\dot{a}^2 + \dot{m_a}m_a a^2,
\end{equation}
where $\rho_a (= \dot{a}^2/2 + m_a^2 a^2/2)$ is the axion energy density.
By averaging the above equation over an oscillation  
($\langle \dot{a} \rangle = \rho_a$ and 
$\langle m_a^2 a^2\rangle = \rho_a$), it is found that
$\rho_a R^3/m_a$ is invariant under adiabatic condition $m_a \gg H$. 
Thus, we obtain the present axion number to entropy ratio as
\begin{equation}
  Y_a^{({\rm cold})}= \frac{n_{a,0}}{s_0} 
  = \beta \left(\frac{\rho_a/m_a}{s}\right)_{T=T_1},
\end{equation}
where $s$ is the entropy density and $s_0$ is its present value.
Here $\beta$ is the 
correction factor taking into account that the adiabatic condition ($m_a \gg H$) 
is not satisfied at the beginning of the oscillation. 
The correction factor was calculated by~\cite{Bae:2008ue} which gives $\beta =1.85$. 
Thus, the present axion density is given by~\cite{Turner:1985si} 
\begin{equation}
  \Omega_{a} h^2 = 0.18\  \theta_1^2\left(\frac{F_a}{10^{12}{\rm GeV}}\right)^{1.19}
  \left(\frac{\Lambda}{400{\rm MeV}}\right),
  \label{eq:density_coherent}
\end{equation}
where $h$ is the present Hubble parameter in units of $100$km/s/Mpc.
Here $\theta_1 = a_1/\eta$ is the initial angle at onset of oscillation. 
When the PQ symmetry is spontaneously broken after inflation, 
$\theta_1$ is random in space and hence we should replace $\theta_1^2$  
by its spatial average, i.e. $\langle \theta_1^2\rangle = \pi^2/3 \times c_{\rm anh}$,
where $c_{\rm anh} (\simeq 2$) is the anharmonic 
correction~\cite{Turner:1985si,Lyth:1991ub}. 
On the other hand, if PQ symmetry is broken before or during inflation, $\theta_1$ 
takes the same value in the whole observable Universe. 
Then, $\theta_1$ is considered as a free parameter. 

The density of the coherent axion oscillation cannot exceed the present 
DM density of the Universe determined from the observations of 
cosmic microwave background (CMB), $\Omega_{\rm CDM} h^2 =0.11$.
This gives the following upper bound on the axion decay constant:
\begin{equation}
    F_a ~\lesssim~  1.4\times 10^{11}~{\rm GeV},
\end{equation}
when the PQ symmetry is broken after inflation. 
For the case of PQ symmetry breaking before or during inflation, 
see Section~\ref{sec:PQ_breaking_after}.\footnote{
	Notice that Equation~\ref{eq:density_coherent} assumes no late-time entropy production
	after the QCD phase transition.
	If there is a late-time entropy production by decaying particles, the abundance is reduced
	and upper bound on the PQ scale is relaxed~\cite{Kawasaki:1995vt}.
}

%%%%%%%%%%%%%%%%%%%%%%%%%%%%%%%%%%%%%
\subsubsection{Hot axion}
\label{sec:hot_axion}
%%%%%%%%%%%%%%%%%%%%%%%%%%%%%%%%%%%%%

Axions are also produced in high-temperature plasma \cite{Masso:2002np,Graf:2010tv}.
The abundance of such hot axions in the KSVZ model, in terms of the number-to-entropy ratio $Y_a\equiv n_a/s$,
was estimated recently in \cite{Graf:2010tv} :
\begin{equation}
	Y_a^{\rm (hot)} \simeq 1.9\times 10^{-3} g_s^6 \ln\left( \frac{1.501}{g_s} \right)
	\left( \frac{10^{12}\,{\rm GeV}}{F_a} \right)^2\left( \frac{T_{\rm R}}{10^{10}\,{\rm GeV}} \right),
\end{equation}
where $T_{\rm R}$ denotes the reheating temperature after inflation.
Notice that the above expression applies for $T_{\rm R} < T_{\rm D}$ with
\begin{equation}
	T_{\rm D} \simeq 9.6\times 10^{6}\,{\rm GeV}\left( \frac{F_a}{10^{10}\,{\rm GeV}} \right)^{2.246}.
\end{equation}
Otherwise, axions are thermalized at $T > T_{\rm D}$ and decouple at $T \simeq T_{\rm D}$.
Then the relic abundance is given by $Y_a^{\rm (hot)}=0.28/g_*(T_{\rm D})$ $(\simeq 2.6\times 10^{-3}$
for $g_*(T_{\rm D}) = 106.75$).

The above calculations were carried out in quark-gluon plasma.
For small $F_a (\lesssim 10^7\,{\rm GeV})$, the axion decoupling may occur after the QCD phase transition.
The decoupling temperature and the resultant axion abundance in such a case were
estimated in \cite{Chang:1993gm,Hannestad:2005df} by taking account of the axion-pion interaction.
Since the axion mass is given by $m_a \gtrsim 0.6$\,eV for $F_a \lesssim 10^7$\,GeV, it contributes to the hot dark matter
and such a contribution is restricted from cosmological observations.
According to recent results \cite{Hannestad:2010yi}, the constraint reads $m_a < 0.91$\,eV assuming massless neutrinos
and $m_a < 0.72$\,eV after marginalizing over the neutrino mass.
This closes a hadronic axion window (see Section~\ref{sec:astro}).

%%%%%%%%%%%%%%%%%%%%%%%%%%%%%%%%%%%%%
\subsection{PQ symmetry breaking after inflation}
\label{sec:PQ_breaking_after}
%%%%%%%%%%%%%%%%%%%%%%%%%%%%%%%%%%%%%

Cosmological consequences of axions are different depending on whether the PQ symmetry 
is broken after inflation or not. 
When the symmetry breaking takes place after inflation, topological defects like  
strings and domain walls are formed in the course of evolution of the PQ scalar
as shown in Section~\ref{sec:PQ-filed_evolution}. 
On the other hand, if the PQ symmetry is broken before or during inflation, 
the produced strings are diluted away and the field value of the axion is the same
in the whole observable Universe. 
Thus, the axion settles into the same minimum when it acquire the mass at the QCD scale,
and hence domain walls are not formed.  
So no domain wall problem exists.
However, in this case,  the axion (which already exists during inflation) obtains 
large fluctuations and produces isocurvature density perturbations which are stringently 
constrained by the CMB observations.  
In this section, we first consider the case where the PQ symmetry breaking occurs 
after inflation and see cosmological consequences the axionic strings and domain walls. 

%%%%%%%%%%%%%%%%%%%%%%%%%%%%%%%%%%%%%
\subsubsection{Axionic strings}
\label{sec:axionic_string}
%%%%%%%%%%%%%%%%%%%%%%%%%%%%%%%%%%%%

\begin{figure}[!tb]
\begin{center}
  \scalebox{1.0}{\includegraphics{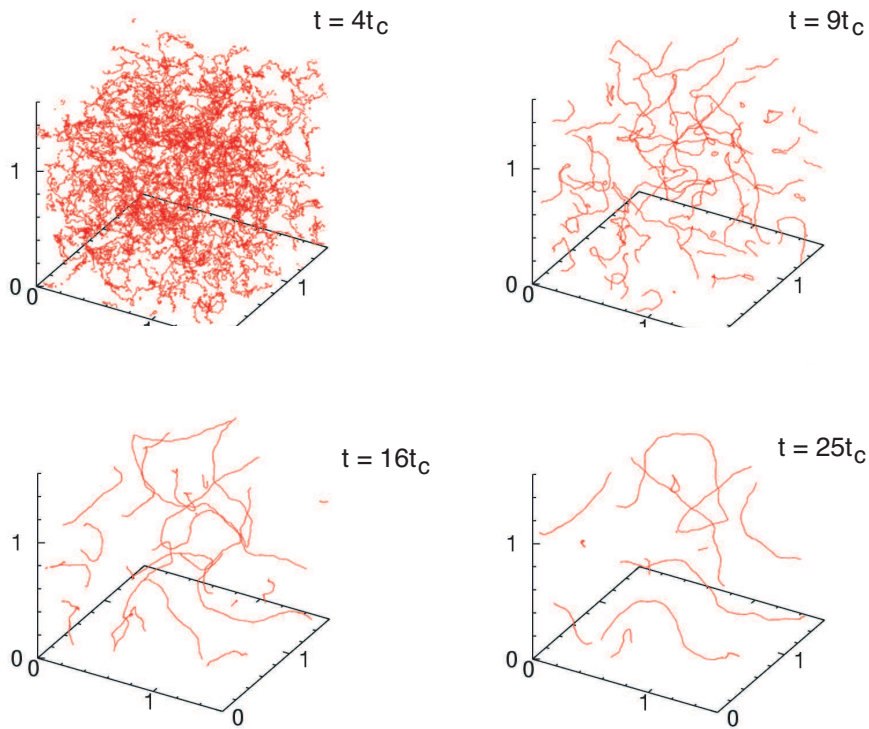}}
  \caption{Evolution of the axionic string networks 
  from the simulation~\cite{Hiramatsu:2010yu}. 
  The panels show the time slices at $t= 4t_c$, $9t_c$, $16t_c$ and $25t_c$ where 
  $t_c$ is the cosmic time corresponding to the critical temperature $T_c$.
  The spatial scale shows a comoving length
  in unit of the horizon size at $t_\mathrm{end}=25t_\mathrm{c}$.
  }
\label{figure2}
\end{center}
\end{figure}

\begin{figure}[tb]
\begin{center}
 \scalebox{1.3}{\includegraphics{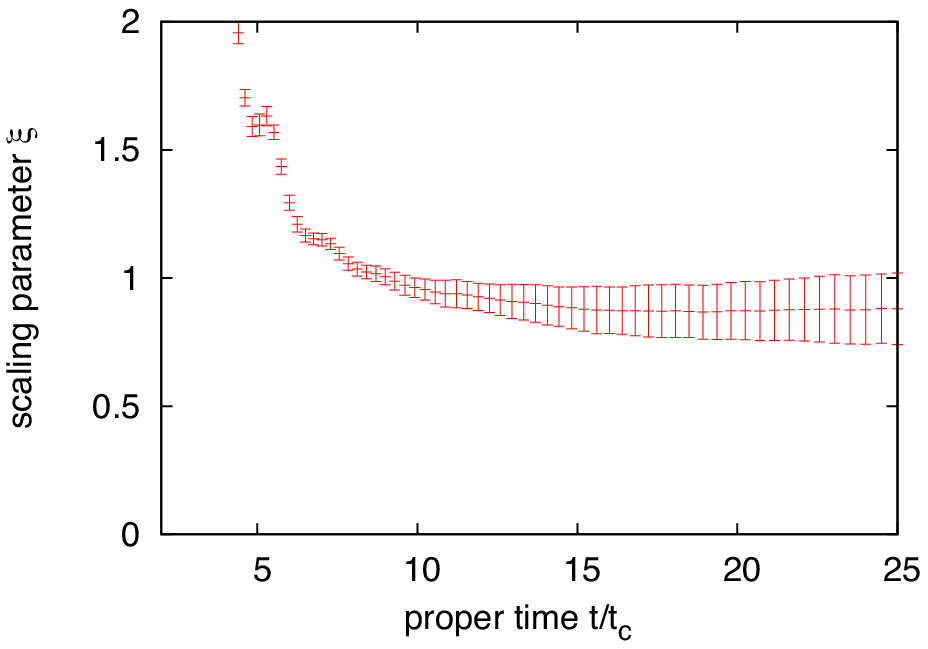}}
  \caption{Time evolution of the length parameter $\xi$ 
  from the simulation~\cite{Hiramatsu:2010yu}.
  }
\label{figure3}
\end{center}
\end{figure}

When the $U(1)_{\rm PQ}$ symmetry is spontaneously broken, one-dimensional 
topological defects called axionic strings are produced. 
Since the $U(1)_{\rm PQ}$ is a global symmetry, the axionic string is a global one.
Unlike local strings which eventually lose their energy by emitting gravitational waves, 
the emission of the Nambu-Goldstone bosons, i.e. axions is a dominant energy loss
process for axionic strings. 
After formation the axionic string networks evolve by emitting axions 
and follow a scaling solution in which
the energy density of the string networks
$\rho_{\rm str}$ is written as   
\begin{equation}
   \rho_{\rm str} = \xi \frac{\mu}{t^2},
\end{equation}
where $\xi$ is the length parameter which is constant in the scaling regime 
and $\mu$ is the string tension given by 
\begin{equation}
   \mu = \pi F_a^2 \ln\left(\frac{t/\sqrt{\xi}}{\delta_s}\right),
\end{equation}
where $\delta_s= 1/\sqrt{\lambda}\eta$ is the width of the strings.
$\xi$ represents average number of infinite strings in a volume 
$t^3$($\sim$ horizon volume) and is determined by numerical 
simulations~\cite{Yamaguchi:1998iv,Yamaguchi:1998gx,Yamaguchi:1999dy,Hiramatsu:2010yu}.
Figure~\ref{figure2} shows the recent 
simulation~\cite{Hiramatsu:2010yu} and the evolution of the length parameter $\xi$
is shown in Figure~\ref{figure3}.
For the axionic string networks, $\xi \simeq 0.7-1.0$.

%%%%%%%%%%%%%%%%%%%% Figure %%%%%%%%%%%%%%%%%%%%%%%%%%%%%%%%%%%%%%
%%%%%%%%%%%%%%%%%%%%%%%%%%%%%%%%%%%%%%%%%%%%%%%%%%%%%%%%%%%%%%%%%%%%
%%%%%%%%%%%%%%%%%  Figure %%%%%%%%%%%%%%%%%%%%%%%%%%%%%%%%%%%%%%%
%%%%%%%%%%%%%%%%%%%%%%%%%%%%%%%%%%%%%%%%%%%%%%%%%%%%%%%%%%%%%%%%%%%

\begin{figure}[tb]
\begin{center}
 \scalebox{1.5}{\includegraphics{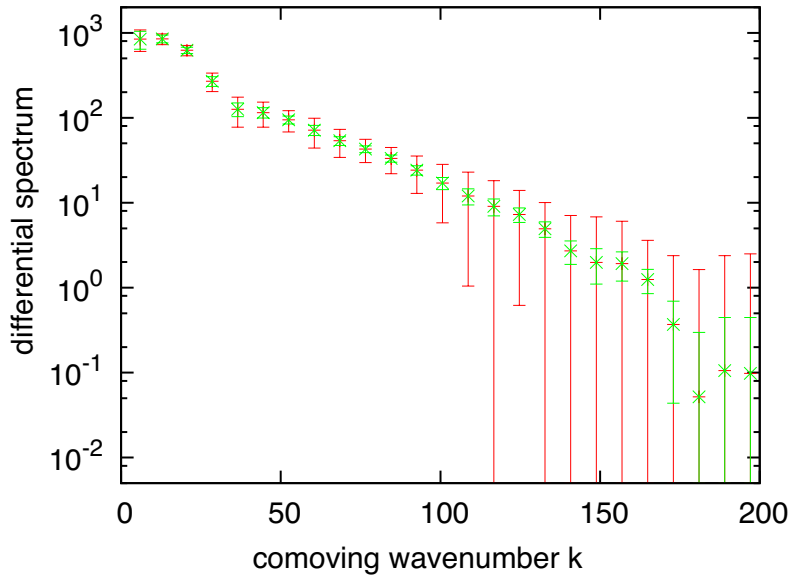}}
  \caption{Differential energy spectrum of radiated axions between 
  $t_1=12.25t_\mathrm{c}$ and $t_2=25t_\mathrm{c}$.   
  Green (red) bars correspond
  to statistical errors alone (statistical and systematic errors).
  The comoving wavenumber $k$ is in units of $(2t_c)^{-1}$.
  Note that the horizon scale corresponds to $k \sim 3$.
  The scale in $y$-axis is arbitrary.}
\label{figure4}
\end{center}
\end{figure}

The axions emitted from the axionic string networks give a significant 
contribution to the present matter density of the Universe.
This was first pointed out by Davis~\cite{Davis:1986xc} 
[see also~\cite{Davis:1989nj,Dabholkar:1989ju}].
The axion density due to the axionic strings depends on the energy spectrum of the 
emitted axions. 
Davis and co-workers claimed that the energy spectrum has a sharp peak at the horizon 
scale~\cite{Davis:1986xc,Davis:1989nj,Dabholkar:1989ju}.
On the other hand, Sikivie and co-workers~\cite{Harari:1987ht,Hagmann:1990mj,Hagmann:2000ja} 
insisted that the spectrum is proportional to $1/k$ ($k$: axion momentum).  
Because the present axion density is given by $\rho_a = m_a(0)n_a$, it is crucial 
how many axions are produced from the strings. 
Since smaller number of axions are emitted for the $1/k$ spectrum, the present density 
of the axions is less important than that for Davis's spectrum. 
This controversy was solved by field theoretical lattice 
simulations~\cite{Yamaguchi:1998gx,Hiramatsu:2010yu} which showed that  
the spectrum is sharply peaked around the horizon scale and exponentially 
suppressed at higher momenta as seen in Figure~\ref{figure4}.
Using this spectrum, the mean reciprocal comoving momentum of the emitted axions
$\langle 1/k\rangle$ which is important in calculating the axion number density is 
estimated as~\cite{Hiramatsu:2010yu} 
\begin{equation}
   \langle 1/k\rangle R(t)  \simeq 0.23 ~\frac{t}{2\pi},
\end{equation}
where $(2\pi)/t$ is the momentum corresponding to the horizon scale. 
Thus, the present axion density due to the axionic strings 
is estimated as
\begin{equation}
   \Omega_{a,{\rm str}}h^2  = 2.0~\xi\left(\frac{F_a}{10^{12}{\rm GeV}}\right)^{1.19}
   \left(\frac{\Lambda}{400{\rm MeV}}\right).
   \label{eq:density_string}
\end{equation}
Notice that the string contribution is larger than the coherent oscillation given 
in Equation~\ref{eq:density_coherent}.

%%%%%%%%%%%%%%%%%% Figure %%%%%%%%%%%%%%%%%%%%%%%%%%%%%%%%
%%%%%%%%%%%%%%%%%%%%%%%%%%%%%%%%%%%%%%%%%%%%%%%%%%%%%%%%%%%%%%%%%%%

%%%%%%%%%%%%%%%%%%%%%%%%%%%%%%%%%%%%%
\subsubsection{Axionic domain walls with $N_{\rm DW}=1$}
\label{sec:axionic_wall}
%%%%%%%%%%%%%%%%%%%%%%%%%%%%%%%%%%%%

Around the QCD scale, the axion acquires mass through QCD instanton effect
and the axion potential is given by Equation~\ref{eq:axion-pot}.
Since the potential has $N_{\rm DW}$ discrete minima and the axion settles down to 
one of the minima, the Universe is divided into many domains with different vacua and domain
walls are formed separating those domains. 
Around an axionic string, the phase of the PQ scalar ($\theta_a= a/\eta$) rotates 
$2\pi$, so $N_{\rm DW}$ domain walls attach each string. 
The cosmological evolutions of the string-domain wall networks are quite different 
between $N_{\rm DW} =1$ and $N_{\rm DW}\ge 2$. 

\begin{figure}[htbp]
  \begin{center}
  \begin{tabular}{c c}
 \resizebox{70mm}{!}{\includegraphics[angle=0]{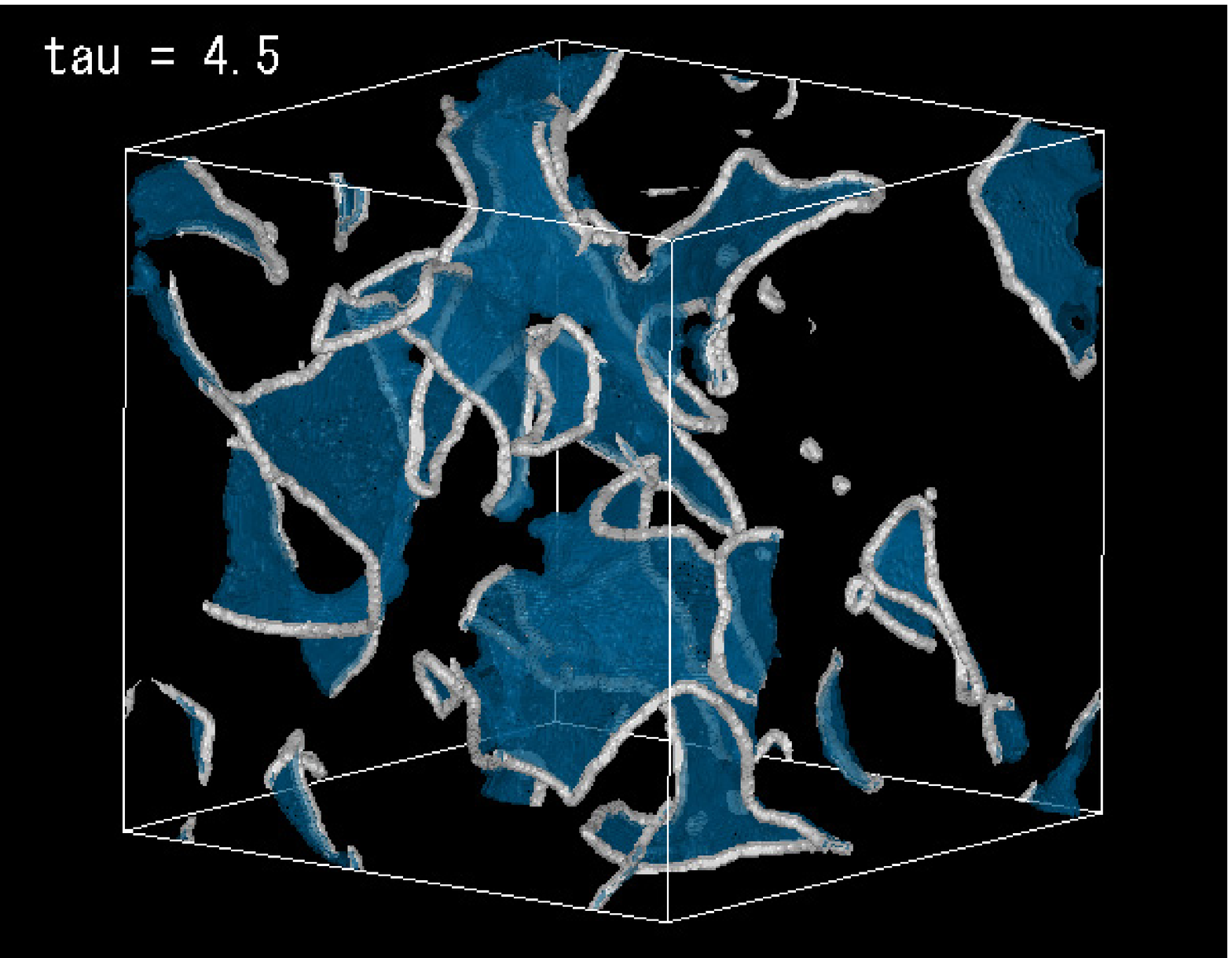}} &
 \resizebox{70mm}{!}{\includegraphics[angle=0]{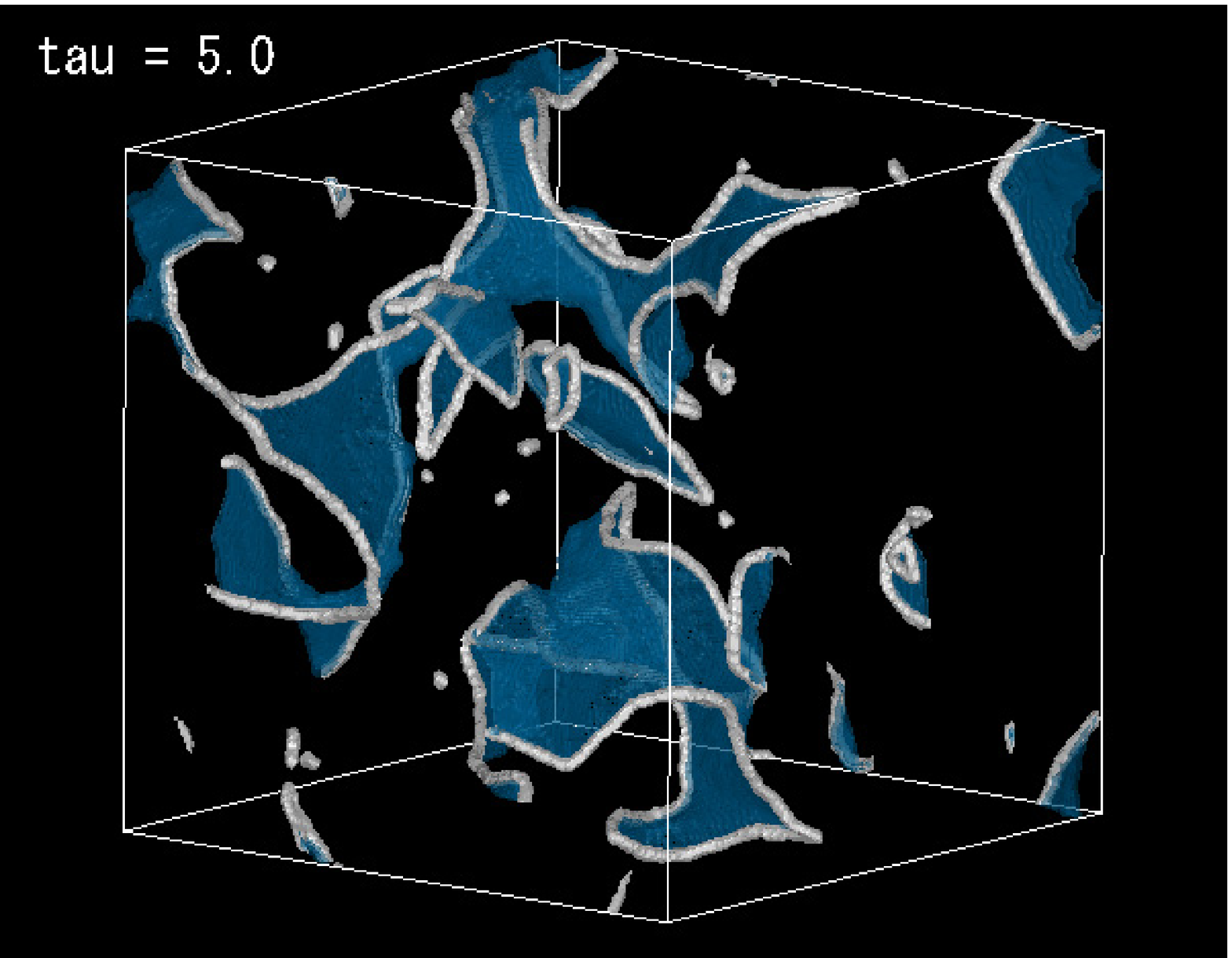}} \\
 \resizebox{70mm}{!}{\includegraphics[angle=0]{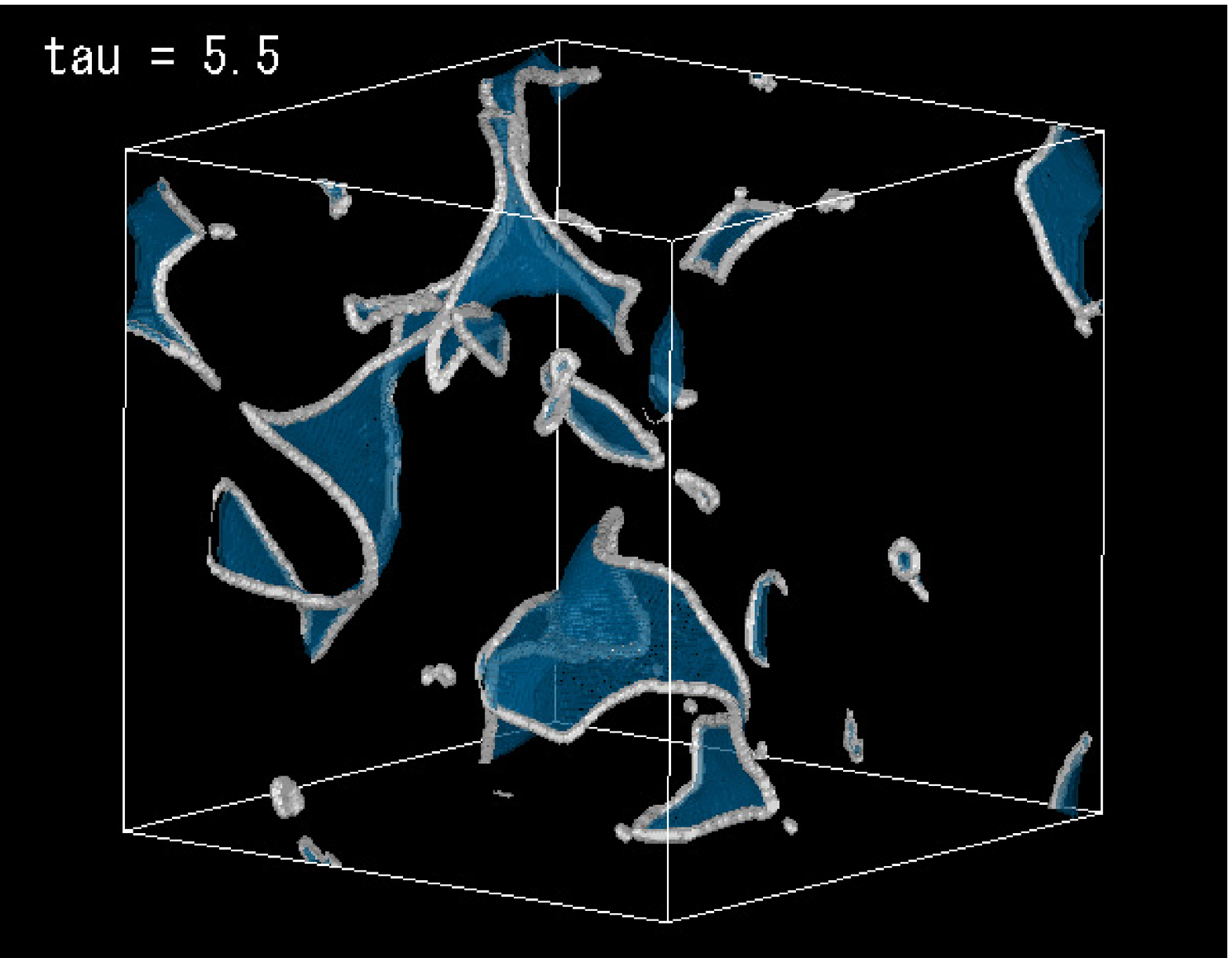}} &
 \resizebox{70mm}{!}{\includegraphics[angle=0]{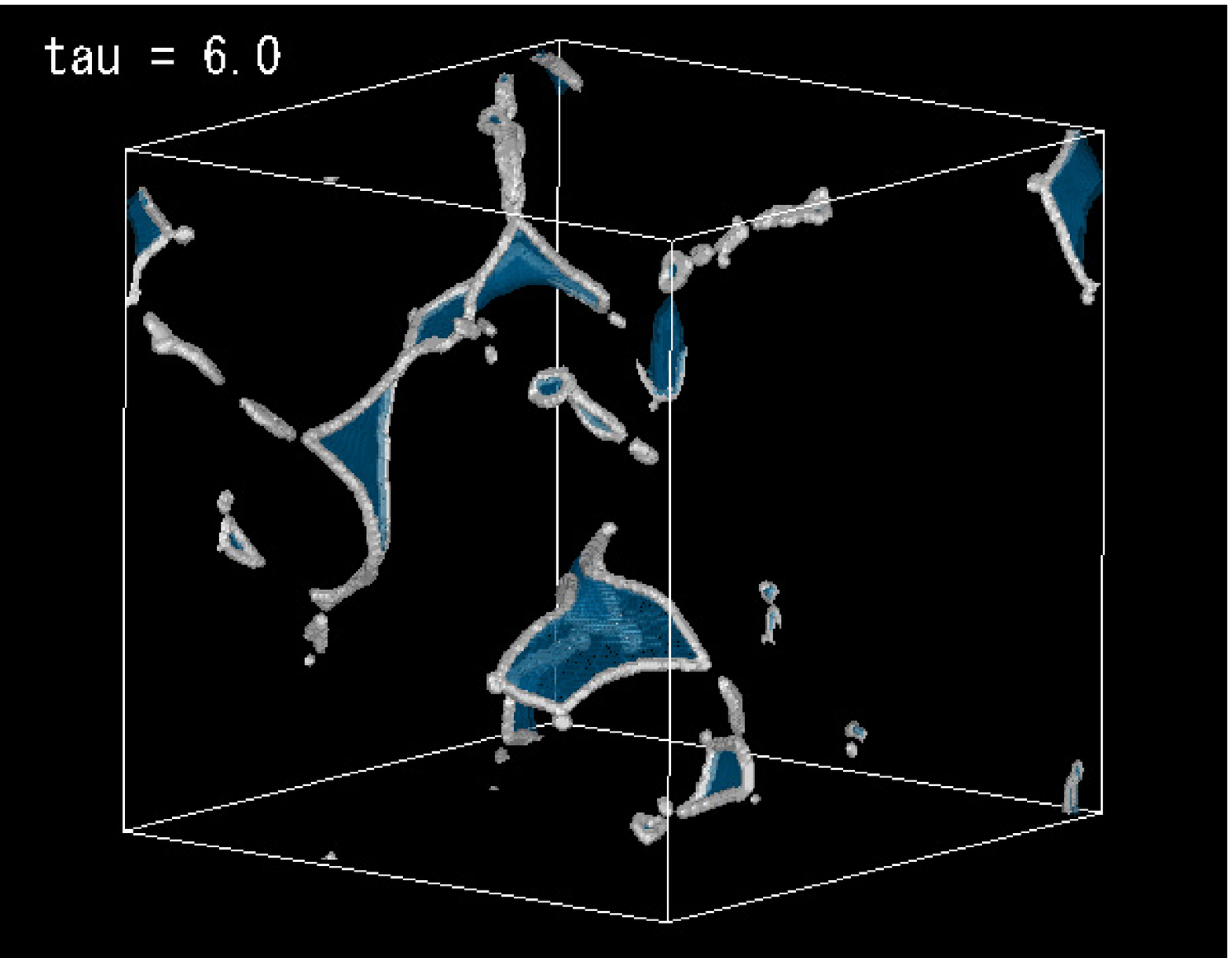}} \\
 \resizebox{70mm}{!}{\includegraphics[angle=0]{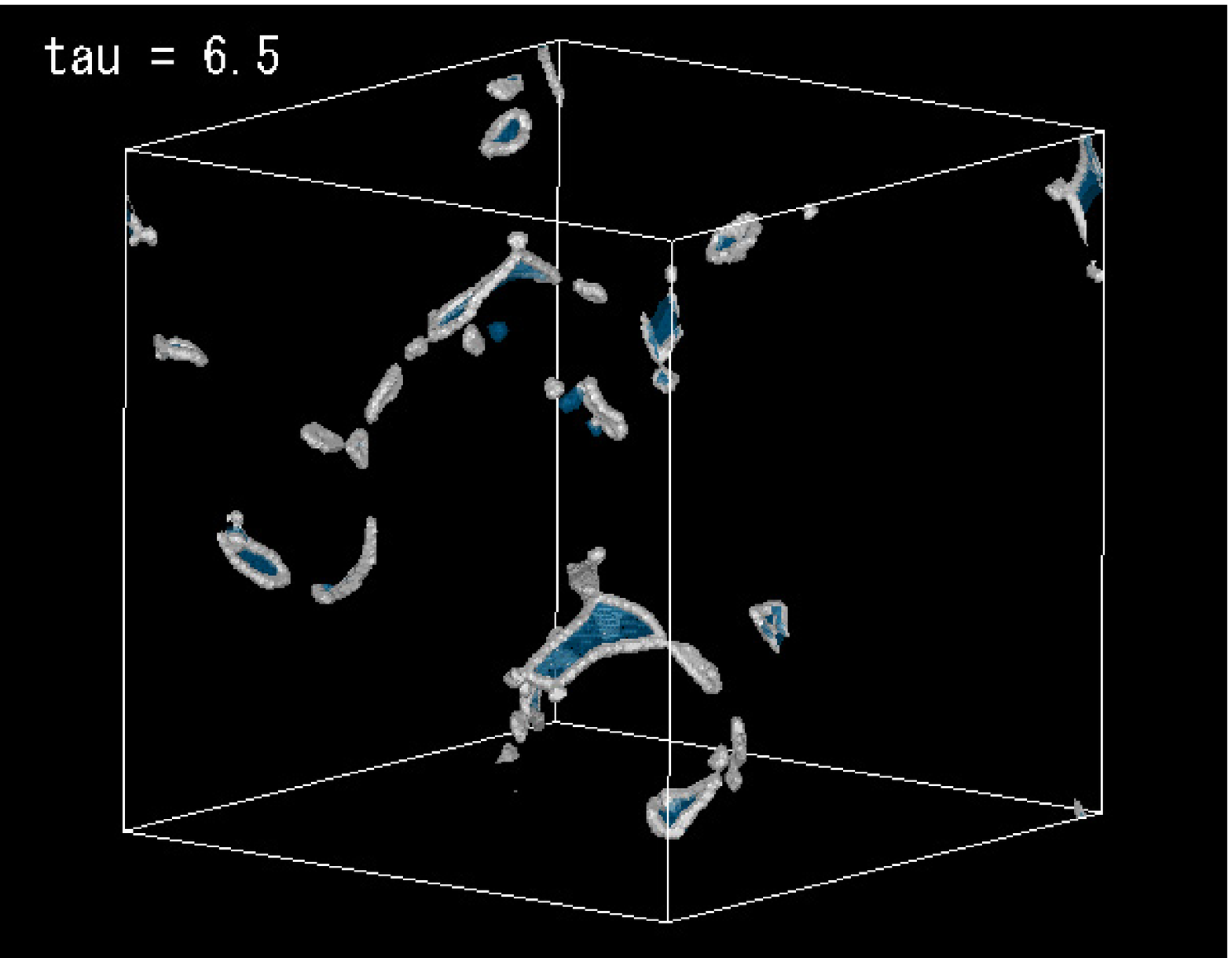}} &
 \resizebox{70mm}{!}{\includegraphics[angle=0]{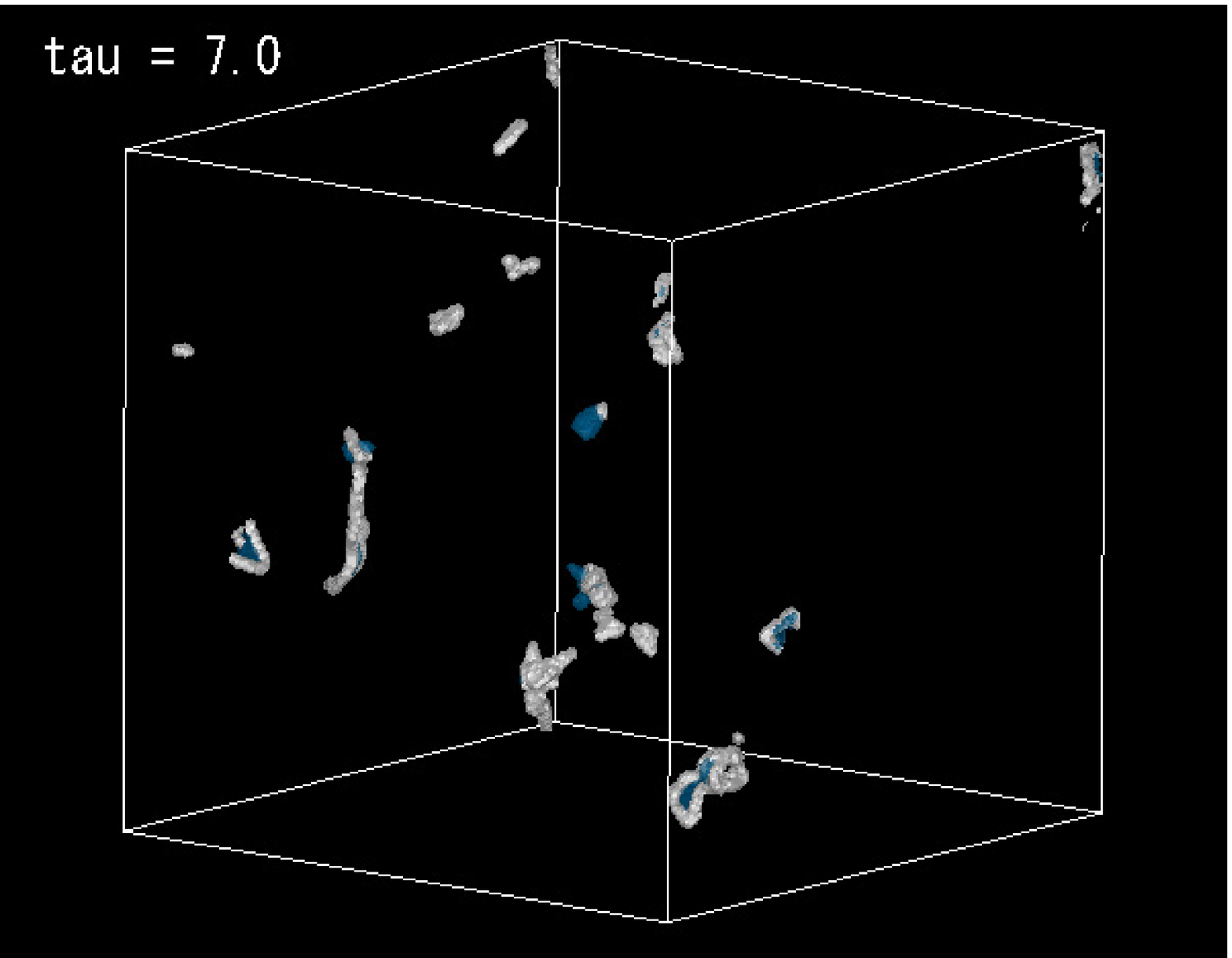}} \\
  \end{tabular}
  \end{center}
  \caption{Evolution of the string-domain wall networks for $N_{\rm DW} =1$.
  The white lines correspond to the position of strings, 
  while the blue surfaces correspond to the position of the center of domain walls.
  ``tau'' in each panel is the conformal time which is related to 
  the cosmic time as ${\rm tau} = \sqrt{t\eta}$.}
  \label{figure5}
\end{figure}

First, let us consider the case of $N_{\rm DW}=1$.
The domain walls with $N_{\rm DW}=1$ are disk-like objects bounded by strings and
collapse by their tension. 
Thus, the string-domain wall networks are unstable for $N_{\rm DW} =1$ and 
they decay soon after formation. 
This was confirmed in 2-dimensional~\cite{Hiramatsu:2010yn} 
and 3-dimensional~\cite{Hiramatsu:2012gg} numerical 
simulations.
In Figure~\ref{figure5} the evolution of the domain walls is shown and 
it is seen that the string-domain wall networks decay and disappear in the Universe. 
So in this case there is no domain wall problem. 

However, a large number of axions are produced in the decay of the string-domain wall
networks and they give a significant contribution to the present matter density
of the Universe. 
Similarly to axions from the axionic strings, there has been controversy 
about the spectrum of the axions emitted from domain walls.
In~\cite{Nagasawa:1994qu} it was found that the spectrum has a peak around $m_a$
which is the width of the domain walls, 
so axions from domain walls are mildly relativistic. 
On the other hand, the authors in~\cite{Chang:1998tb,Hagmann:1998me} 
claimed that the mean axion energy
is larger than that in~\cite{Nagasawa:1994qu} by a factor $20$.
This conclusion comes from the reasoning that the energy of the domain walls 
is converted into strings which emits axion with spectrum proportional to $1/k$ 
following~\cite{Harari:1987ht,Hagmann:1990mj,Hagmann:2000ja}.  
The recent 3-dimensional lattice simulation~\cite{Hiramatsu:2012gg} 
settled this problem and showed that 
the axions produced in the string-wall decay are mildly relativistic with mean 
momentum $\langle k \rangle /R \simeq 3m_a$ which is consistent 
with~\cite{Nagasawa:1994qu}.
The energy density of the axions today is then estimated as
\begin{equation}
   \Omega_{a,{\rm wall}}h^2 = (5.8\pm 2.8)~\left(\frac{F_a}{10^{12}{\rm GeV}}\right)^{1.19}
   \left(\frac{\Lambda}{400{\rm MeV}}\right).
\end{equation}
Comparing to the contributions from the coherent oscillation 
(Equation~\ref{eq:density_coherent} )
and strings (Equation~\ref{eq:density_string}), it is found that the axions from the string-wall
networks gives a dominate contribution to the present DM density. 

%%%%%%%%%%%%%%%%%%%%%%%%%% Figure %%%%%%%%%%%%%%%%%%%%%%%%%%%%%%%

%%%%%%%%%%%%%%%%%%%%%%%%%%%%%%%%%%%%%%%%%%%%%%%%%%%%%%%%%%%%%%%%%%%%%

The total density of the cold axions is given by 
\begin{equation}
   \Omega_{a,{\rm tot}}h^2 = (8.4\pm 3.0)~\left(\frac{F_a}{10^{12}{\rm GeV}}\right)^{1.19}
   \left(\frac{\Lambda}{400{\rm MeV}}\right),
\end{equation}
where we take $\xi = 1.0\pm 0.5$ in Equation~\ref{eq:density_string}.
Thus we obtain the following constraint on the axion decay constant:
\begin{equation}
    F_a ~\lesssim~  (2.0-3.8)\times 10^{10}~{\rm GeV}.
\end{equation}
%%

%%%%%%%%%%%%%%%%%%%%%%%%%%%%%%%%%%%%%
\subsubsection{Axionic domain walls with $N_{\rm DW}\ge 2$}
\label{sec:axionic_wall_N}
%%%%%%%%%%%%%%%%%%%%%%%%%%%%%%%%%%%%

The wall-string networks with $N_{\rm DW} \ge 2$ have complicated structures 
and do not decay contrary to the walls with $N_{\rm DW}=1$. 
After formation, the long-lived wall-string networks evolve into the scaling regime
as shown in the simulations~\cite{Ryden:1989vj,Hiramatsu:2010yn}.
The domain wall density is written as 
\begin{equation}
  \rho_{\rm wall} = {\cal A} \frac{\sigma_{\rm wall}}{t},
\end{equation}
where $\sigma_{\rm wall}$ is the wall tension given by 
$\sigma_{\rm wall} =9.23m_a F_a^2$.
Here ${\cal A}$ is the surface parameter which becomes constant in the scaling regime. 
Since the total cosmic density decreases as $1/t^2$, the domain walls soon dominate
the Universe, which conflicts the standard cosmology. 

To solve the domain wall problem, one can introduce an additional term 
in the axion potential which explicitly breaks the $Z_{N_{\rm DW}}$ symmetry, 
\begin{equation}
  \delta V = -\Xi \eta^3 (\phi e^{i\delta} + {\rm h.c.}),
\end{equation}
where $\Xi$ is a small parameter describing the size of the explicit $Z_{N_{\rm DW}}$
breaking and $\delta$ is the phase. 
This term is called a ``bias'' and lifts the degenerated vacua so that the potential
has a unique minimum at $a\simeq 0$.
The energy difference between the neighboring vacuum produces a volume pressure and
makes the domain walls accelerate toward the false vacuum. 
Thus, the false vacuum regions shrink and the domain walls annihilate each other.
Let us estimate the decay epoch of the string-wall networks with bias.
The pressure due to the bias term is roughly given by the difference of 
neighboring vacua which is $p_V \sim 4\pi\Xi\eta^4/N_{\rm DW}$.
On the other hand,  the pressure from the surface tension which straightens 
the wall up to the horizon scale is $p_T \sim \sigma_{\rm Wall}/t$.
The domain walls collapse when $p_V \sim p_T$, so the decay time $t_{\rm dec}$ is
estimated as
\begin{equation}
   t_{\rm dec} \sim \frac{\sigma_{\rm wall}}{\Xi \eta^4/N_{\rm DW}}.
\end{equation}
Using $\sigma_{\rm wall} \sim m_a \eta^2/N_{\rm DW}^2$ we can write the above equation as
\begin{equation}
   t_{\rm dec} = \alpha  \frac{m_a}{\Xi \eta^2 N_{\rm DW}},
\end{equation}
where $\alpha \simeq  18$ is determined by the numerical simulation~\cite{Hiramatsu:2010yn}.
  
\begin{figure}[htbp]
\begin{center}
\includegraphics[scale=1.0]{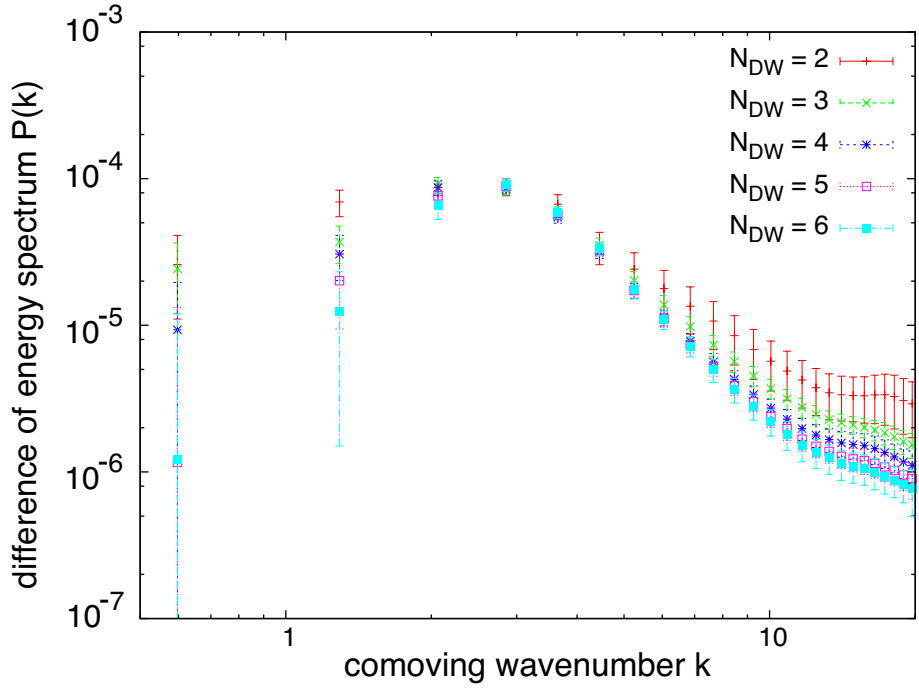}
\end{center}
\caption{The spectra of axions produced by string-wall networks 
for various values of $N_{\mathrm{DW}}$ in the numerical simulation~\cite{Hiramatsu:2012sc}. 
The spectra shown in this figure are the difference of power spectra evaluated 
at two different time steps $t_1 =49\eta^{-1}$ and $t_2=400\eta^{-1}$.
The momentum scale corresponding to the axion mass is given by $k \sim R(t_2) m_a\sim 2$.
}
\label{figure6}
\end{figure}

The string-wall networks continuously emit axions until they decay at $t_{\rm dec}$.
The spectrum of the emitted axions has a peak around the momentum scale 
determined by the mass of axion as shown in Figure~\ref{figure6} 
and the mean momentum of the axions $\epsilon_a$ is estimated as 
$\epsilon_a = \langle k\rangle /R \simeq 3 m_a$.
This is the same as the case for $N_{\rm DW} =1$. 
Then we can obtain the present density of axions from the string-wall networks,
\begin{equation}
   \Omega_{a,{\rm wall}}h^2 \simeq 0.3~ {\cal A} N_{\rm DW}^{-3/2}
   \left(\frac{\Xi}{10^{-52}}\right)^{-1/2} 
   \left(\frac{F_a}{10^{10}{\rm GeV}}\right)^{-1/2}.
   \label{eq:wall_axion_N}
\end{equation}
Since the long-lived walls produce more axions than the wall with $N_{\rm DW}=1$,
the cosmological constraint is more stringent as seen below. 

Gravitational waves are also produced from the string-wall networks. 
Since the characteristic length scale of the domain walls is the Hubble time $\sim t$,
the quadruple moment $Q_{ij}$ of the system is estimated as 
$Q_{ij} \sim \rho_{\rm wall} t^3 t^2\sim \sigma_{\rm wall}{\cal A} t^4$. 
Using the quadrupole formula the energy of the gravitational waves is 
$E_{\rm gw} \sim G{\cal A}^2\sigma_{\rm wall}^2 t^3$. 
Thus we can write the energy density of the produced gravitational waves as 
\begin{equation}
   \rho_{\rm gw} = \epsilon_{\rm gw} G {\cal A}^2\sigma_{\rm wall}^2,
\end{equation}
where $\epsilon_{\rm gw}$ is the efficiency of gravitational weaves 
and $\epsilon_{\rm gw} \simeq 5$ from the simulation~\cite{Hiramatsu:2012sc}.
The present density of the gravitational waves is calculated as 
\begin{equation}
   \Omega_{\rm gw} h^2 \simeq 10^{-19} {\cal A}^2 N_{\rm DW}^{-6}
   \left(\frac{\Xi}{10^{-52}}\right)^{-2} 
   \left(\frac{F_a}{10^{10}{\rm GeV}}\right)^{-4}.
\end{equation}
The amplitude of gravitational waves produced by the domain walls is too small
to be detected in the current and future-planned experiments for allowed values of
$\Xi$ and $F_a$ (see below). 

\begin{figure}[htbp]
\begin{center}
\includegraphics[scale=0.9]{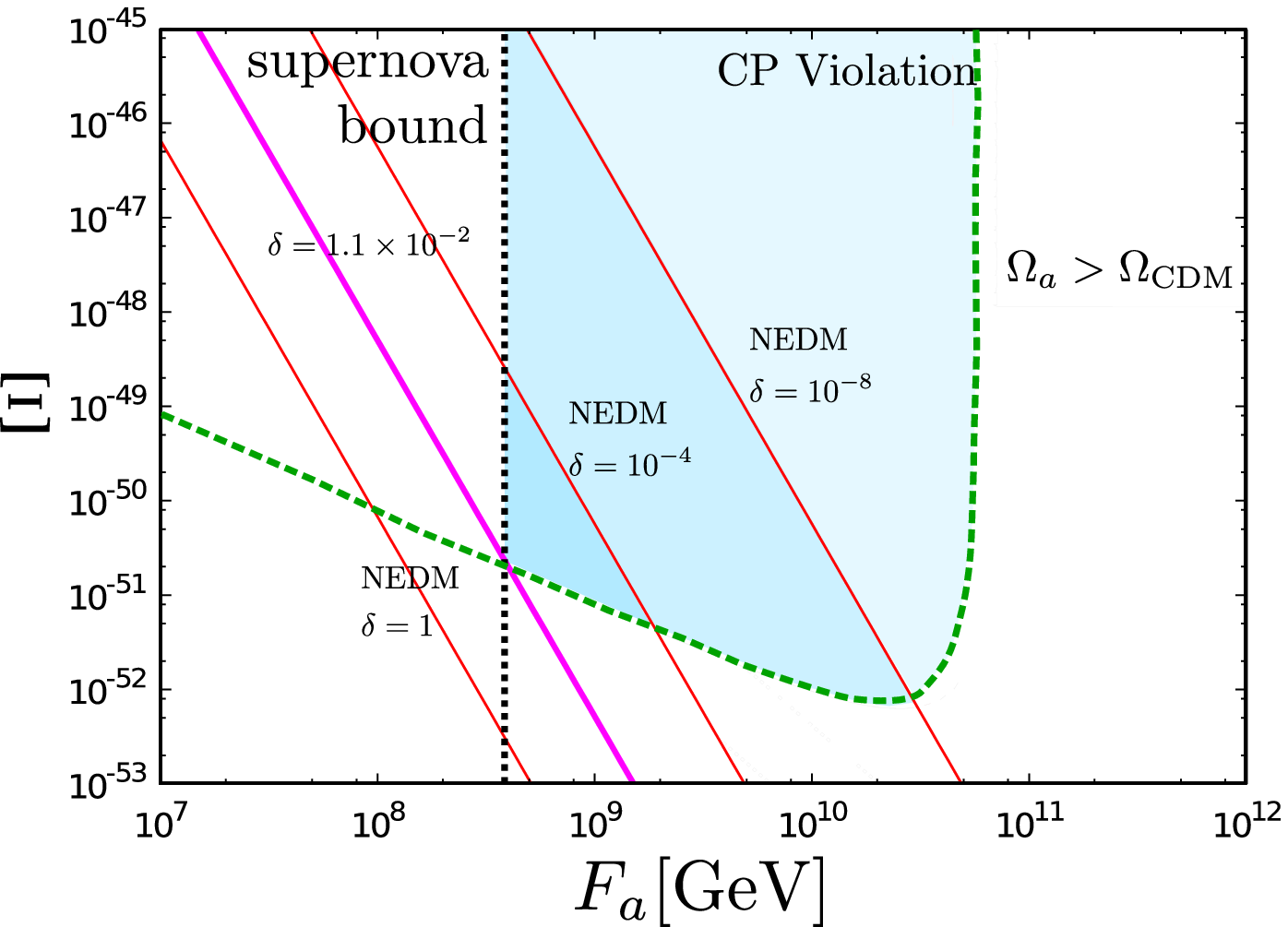}
\end{center}
\caption{The various observational constraints in the parameter space of $F_a$ and $\Xi$.
The green dashed-line represents the parameter region satisfying  
$\Omega_{a,{\rm tot}} < \Omega_{\mathrm{CDM}}$,
and the region below this line is excluded since the relic abundance of axions 
exceeds the cold dark matter abundance observed today.
The vertical dotted-line represents the bound from the observation of supernova 1987A.
The red solid-lines represent the NEDM bound for $\delta=1$, $10^{-4}$, and $10^{-8}$.
The region above these lines is excluded since it leads to an experimentally 
unacceptable amount of CP-violation.
The pink line represents the NEDM bound for the critical value $\delta=1.1\times 10^{-2}$.
There are still allowed regions if the value of $\delta$ is smaller than this critical value.
In this figure, we fixed other parameters as $N_{\mathrm{DW}}=6$, $\epsilon_a=1.5$ 
and ${\cal A}=2.6$.}
\label{figure7}
\end{figure}

Let us examine whether we have a consistent scenario with $N_{\rm DW} \ge2$ or not.
First the cosmic density of the cold axions should not exceed the CDM density 
measured by the CMB observations which gives 
\begin{equation}
   \Omega_{a,{\rm tot}}= \Omega_{a} + \Omega_{a,{\rm str}} + \Omega_{a,{\rm wall}} < 0.11.
\end{equation}
Second, the bias term shifts the minimum of the axion potential by
\begin{equation}
   \bar{\theta} = \frac{\langle a\rangle}{\eta} = 
   \frac{2\Xi N_{\rm DW}^3 F_a^2 \sin\delta}{m_a^2 + 2\Xi N_{\rm DW}^2 F_a^2\cos\delta},
   \label{eq:theta_shift}
\end{equation}
which violates CP. 
From the experimental upper bound on NEDM~\cite{Baker:2006ts}, $\theta$
should satisfy
\begin{equation}
   \bar{\theta} < 0.7\times 10^{-11},
\end{equation}
which leads to the constraints on $\Xi$ and $F_a$ through Equation~\ref{eq:theta_shift}.
Finally, we have the astrophysical constraint from SN1987A, $F_a > 4\times 10^8$~GeV.
In Figure~\ref{figure7} these constraints are shown. 
If the phase of the bias term $\delta$ is $O(1)$, there is no allowed region 
and the allowed region appears for $\delta \lesssim 10^{-2}$. 
Therefore, when the PQ symmetry is broken after inflation,  
the axion models with $N_{\rm DW} \ge2$ are excluded unless the phase 
of the bias term is fine-tuned.

%%%%%%%%%%%%%%%%%%%%%%%%%%%% Figure %%%%%%%%%%%%%%%%%%%%%%%%%%%%%%%%%%%%%%
%%%%%%%%%%%%%%%%%%%%%%%%%%%%%%%%%%%%%%%%%%%%%%%%%%%%%%%%%%%%%%%%%%%%%%%%%%%%%%%

%%%%%%%%%%%%%%%%%%%%%%%%%%%% Figure %%%%%%%%%%%%%%%%%%%%%%%%%%%%%%%%%%%%%%
%%%%%%%%%%%%%%%%%%%%%%%%%%%%%%%%%%%%%%%%%%%%%%%%%%%%%%%%%%%%%%%%%%%%%%%%%%%%%%%%%

%%%%%%%%%%%%%%%%%%%%%%%%%%%%%%%%%%%%%
\subsection{PQ symmetry breaking before/during inflation}
\label{sec:PQ_breaking_before_inf}
%%%%%%%%%%%%%%%%%%%%%%%%%%%%%%%%%%%%%

If the PQ symmetry is broken before or during inflation,  inflation expands 
a domain with some PQ phase  $\theta_a$ to the size larger 
than the present observable Universe.
Thus, there are no topological defects and hence no axions from the axionic domain
walls and strings in the Universe. 
The present axion density comes from the coherent oscillation and is given by Equation~\ref{eq:density_coherent} :
\begin{equation}
  \Omega_{a} h^2 = 0.18\  \theta_a^2\left(\frac{F_a}{10^{12}{\rm GeV}}\right)^{1.19}
  \left(\frac{\Lambda}{400{\rm MeV}}\right),
  \label{eq:coherent_density_inf}
\end{equation}
where $\theta_a$ is constant in the whole observable Universe and considered as a free 
parameter. 
This leads to the upper bound on the axion decay constant, 
\begin{equation}
   F_a < 6.6\times 10^{11}\ \theta_a^{-1.68}~{\rm GeV} .
   \label{eq:fa_constraint_inf}
\end{equation}
Notice that $F_a$ much larger than $10^{12}$~GeV is allowed for $\theta_a \ll 1$.

During inflation the axion field obtains the fluctuations whose power spectrum is 
given by
\begin{equation}
  {\cal P}_{\delta a} = \frac{H_{\rm inf}^2}{4\pi^2}.
  \label{eq:axion_fluctuation}
\end{equation}
Here the power spectrum ${\cal P}_{\delta a}$ is defined by
\begin{equation}
   \langle \delta a(\vec{k}) \delta a (\vec{k}')\rangle 
   = (2\pi)^3\delta^{(3)}(\vec{k} + \vec{k}')\frac{2\pi^2}{k^3}{\cal P}_{\delta a}(k),
   \label{eq:two-point-corr}
\end{equation}
where $\delta a (\vec{k})$ is the Fourier component of $\delta a(\vec{x})$ and
$\langle \cdots\rangle$ denotes the ensemble average.
Here notice that the fluctuations of the axion field
also give a contribution to the mean energy density. 
Using $\langle \delta a^2(\vec{x})\rangle \simeq H_{\rm inf}^2/4\pi^2$ derived from 
Equations~\ref{eq:axion_fluctuation} and \ref{eq:two-point-corr},  
Equation~\ref{eq:coherent_density_inf} should be 
replaced by
\begin{equation}
  \Omega_{a} h^2 = 0.18 
  \left[ \theta_a^2 +\left(\frac{H_{\rm inf}}{2\pi F_a}\right)^2\right]
  \left(\frac{F_a}{10^{12}{\rm GeV}}\right)^{1.19}
  \left(\frac{\Lambda}{400{\rm MeV}}\right).
\end{equation}
When the axion fluctuations are larger than the homogeneous value $F_a\theta_a$,
the cosmic axion density is dominated by the fluctuations and the constraint on 
$F_a$ and $H_{\rm inf}$ is given by~\cite{Lyth:1991ub} 
\begin{equation}
   H_{\rm inf} < 5.0\times 10^{12}{\rm GeV} 
   \left(\frac{F_a}{10^{12}{\rm GeV}}\right)^{0.41}
   ~~~~~{\rm for}~~~~F_a < \frac{H_{\rm inf}}{2\pi\theta_a}.
\end{equation}
On the other hand, for $F_a > H_{\rm inf}/2\pi\theta_a$, 
Equation~\ref{eq:fa_constraint_inf} is valid.

The axion fluctuations turn into isocurvature density perturbations of the axions 
when the axions acquire mass at the QCD scale~\cite{Axenides:1983hj,
Seckel:1985tj,Linde:1985yf,Linde:1990yj,Turner:1990uz,Linde:1991km,Lyth:1991ub}. 
The definition of the axion isocurvature perturbations $S_a$ is 
\begin{equation}
   S_a(\vec{x}) = 3(\zeta_a -\zeta),
\end{equation}
where $\zeta_a$ and $\zeta$ are the curvature perturbations on the uniform density 
slices of the axion and the total matter, respectively. 
According to $\delta N$ formalism~\cite{Sasaki:1995aw,Lyth:2004gb} 
the axion energy density $\rho_a(\vec{x})$ 
on the uniform density slice of the total matter is given by 
$\rho_a(\vec{x}) = \bar{\rho}_a e^{S_a}$ where $\bar{\rho}_a$ is the mean density 
of the axion. 
Since $\rho_a(\vec{x})\propto  (a_i + \delta a(\vec{x}))^2$ with $a_i = F_a\theta_a$,
\begin{equation}
   e^{S_a} = 1 + 2\frac{\delta a(\vec{x})}{a_*^2} 
   + \frac{\delta a^2(\vec{x})-\langle \delta a^2\rangle}{a_{*}^2},
\end{equation}
where $a_{*}^2 = a_i^2 + \langle \delta a^2\rangle$.
Define $r$ as the ratio of the mean axion density $\bar{\rho}_a$ 
to the total CDM density $\bar{\rho}_{\rm CDM}$, 
the CDM density on the uniform density slice of the total matter is 
\begin{equation}
   \rho_{\rm CDM}(\vec{x}) = \bar{\rho}_{\rm CDM} 
   \left[(1-r) + r e^{S_a}\right]. 
\end{equation}
Thus, we obtain the CDM isocurvature density perturbations $S_{\rm CDM}$ as
\begin{eqnarray}
   S_{\rm CDM} & = & 
   \ln\left(\frac{\rho_{\rm CDM}(\vec{x})}{\bar{\rho}_{\rm CDM}}\right)
   \nonumber \\
   & =& 2r \frac{F_a\theta_a \delta a}{a_{*}^2}
   + r \frac{\delta a^2 - \langle \delta a^2\rangle}{a_{*}^2}.
   \label{eq:iso_CDM}
\end{eqnarray}
The energy fraction of the axion in CDM is given by
\begin{equation}
  r = 1.6 \left[ \theta_a^2 +\left(\frac{H_{\rm inf}}{2\pi F_a}\right)^2\right]
  \left(\frac{F_a}{10^{12}{\rm GeV}}\right)^{1.19}
  \left(\frac{\Lambda}{400{\rm MeV}}\right).
\end{equation}
The power spectrum of the CDM isocurvature perturbations is written as
\begin{eqnarray}
   \langle S_{\rm CDM}(\vec{k}) S_{\rm CDM}(\vec{k}')\rangle 
   & = & (2\pi)^3\delta^{(3)}(\vec{k} + \vec{k}')\frac{2\pi^2}{k^3}{\cal P}_{S_{\rm CDM}}(k),
   \\
   {\cal P}_{S_{\rm CDM}} & = & 4r^2 
   \frac{(H_{\rm inf}/2\pi)^2}{(F_a\theta_a)^2 + (H_{\rm inf}/2\pi)^2},
\end{eqnarray}
The amplitude of the isocurvature perturbations is stringently constrained 
by the CMB observations~\cite{Moodley:2004nz,Beltran:2004uv,Bean:2006qz,Trotta:2006ww,
Keskitalo:2006qv,Kawasaki:2007mb,Sollom:2009vd,Valiviita:2012ub}.
The analysis using WMAP 7 year data~\cite{Komatsu:2010fb} finds
$\alpha = {\cal P}_{S_{\rm CDM}}(k_*)/{\cal P}_\zeta(k_*) < 0.15$ at the reference scale
$k_* = 0.002{\rm Mpc}^{-1}$  where ${\cal P}_\zeta$ is 
the power spectrum of the curvature perturbations.
This limit is translated into the constraint on $F_a$ and $H_{\rm inf}$
as 
\begin{equation}
   \left[ \theta_a^2 +\left(\frac{H_{\rm inf}}{2\pi F_a}\right)^2\right]
   \left(\frac{H_{\rm inf}}{2\pi F_a}\right)^2
   \left(\frac{F_a}{10^{12}{\rm GeV}}\right)^{2.38} < 3.6\times 10^{-11}.
\end{equation}
Here we have used ${\cal P}_\zeta(k_*) = 2.43\times 10^{-9}$. 

Although the axion fluctuation $\delta a$ is Gaussian, $\delta a^2$-term 
in Equation~\ref{eq:iso_CDM} is not Gaussian and hence leads to non-Gaussianity
in the isocurvature perturbations in CDM~\cite{Boubekeur:2005fj,Kawasaki:2008sn,Langlois:2008vk,
Kawasaki:2008pa,Kawakami:2009iu,Langlois:2011zz,Langlois:2010fe,Langlois:2012tm}. 
The non-Gaussianity clearly appears in the bispectrum $B(k_1,k_2,k_3)$ 
of the isocurvature perturbations which is defined as~\cite{Kawasaki:2008sn}
\begin{eqnarray}
   \langle S_{\rm CDM}(\vec{k}_1)S_{\rm CDM}(\vec{k}_2)S_{\rm CDM}(\vec{k}_3)\rangle
     =  (2\pi)^3 B_{S_{\rm CDM}}(k_1,k_2,k_3)  
    \delta^{(3)}(\vec{k}_1+\vec{k}_2+\vec{k}_3),\\
     B_{S_{\rm CDM}}(k_1,k_2,k_3)  = f_S 
   \left(\frac{(2\pi^2)^2}{k_1^3 k_2^3}{\cal P}_{S_{\rm CDM}}(k_1){\cal P}_{S_{\rm CDM}}(k_2) + (2~{\rm perms})\right),
   ~~~~
\end{eqnarray}
where $f_S$ is the non-linearity parameter which is given by $1/(2r)$ for
the isocurvature perturbations produced by the axion fluctuations.\footnote{%
The parameter $f_S$ is related to $f_{\rm NL}^{S,SS}$ in \cite{Langlois:2012tm} 
through the relation $f_{\rm NL}^{S,SS} = \alpha^2 f_S$. 
This is further related to $f_{\rm NL}^{\rm (iso)}$ in \cite{Kawasaki:2008pa} 
through $f_{\rm NL}^{S,SS}/27 = (6/5)f_{\rm NL}^{\rm (iso)}$. 
}

\begin{figure}
  \begin{center}
  \begin{tabular}{cc}
    \hspace{0mm}\scalebox{.7}{\includegraphics{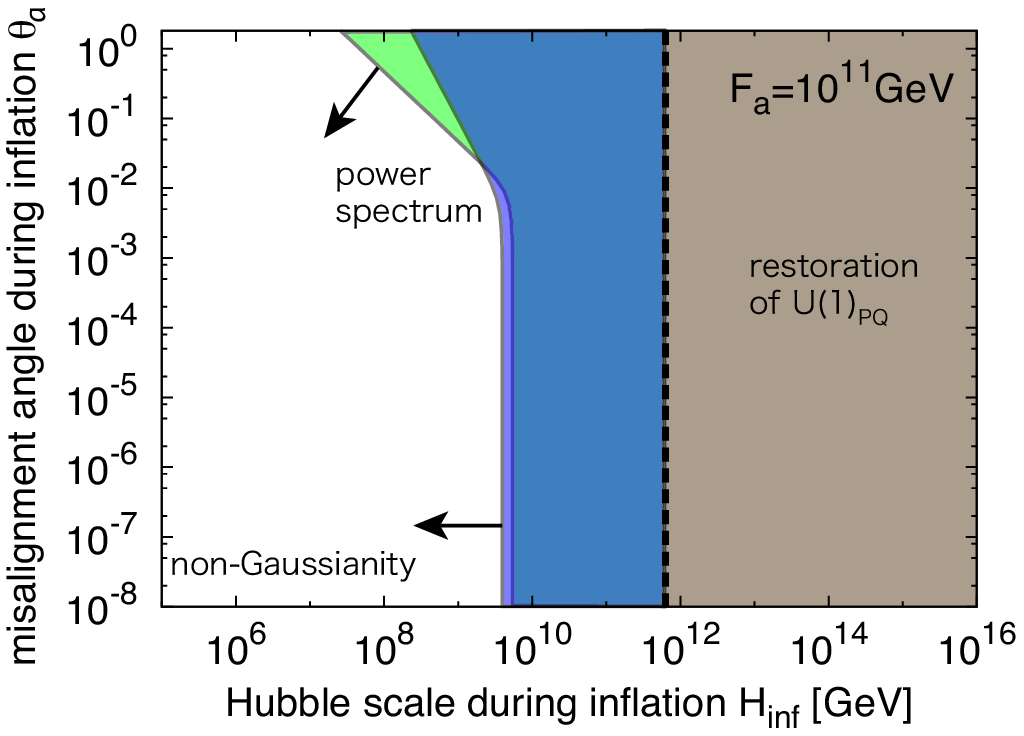}} 
  \hspace{0mm}\scalebox{.7}{\includegraphics{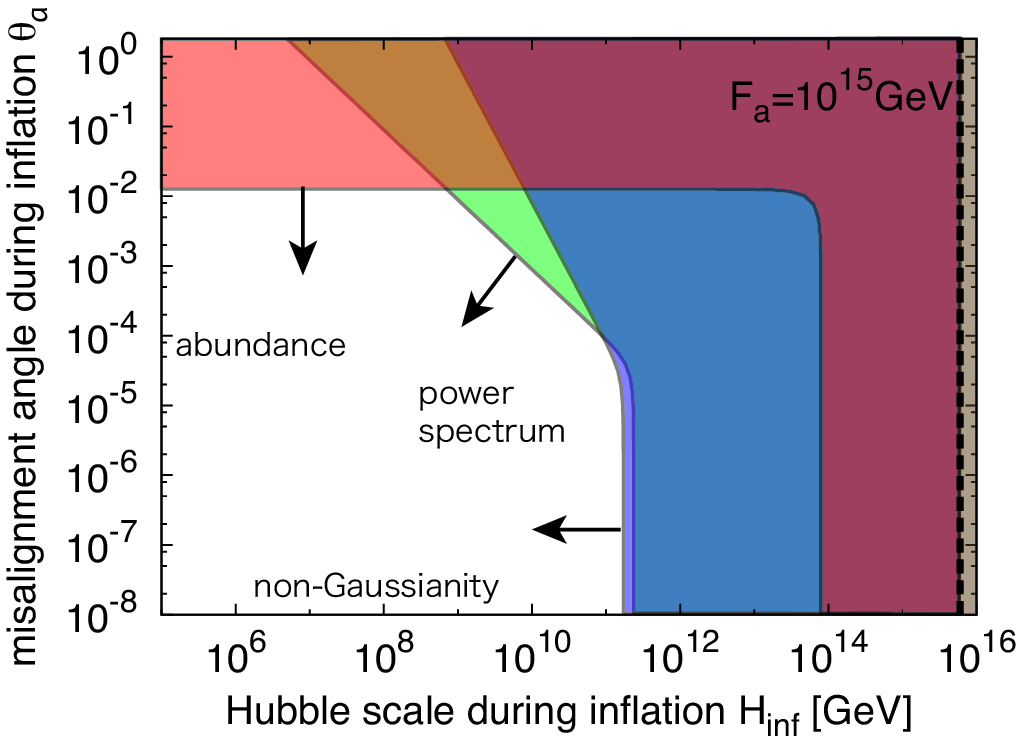}}
  \end{tabular}
  \end{center}
  \caption{
  Constraints on axion isocurvature model in the
  $H_{\rm inf}$-$\theta_a$ plane for $F_a=10^{11}$GeV (left panel) ans 
  $F_a=10^{15}$GeV(right panel).
  Shaded regions are excluded by cosmological considerations. 
  At small $\theta_a$, the constraint on $H_{\rm inf}$ from the non-Gaussianity
  is slightly better than one from the power spectrum.
  When $H_{\rm inf}/2\pi > F_a$ the PQ symmetry restore during inflation.
  }
  \label{figure8}
\end{figure}

At present CMB is the best probe of non-Gaussianity in isocurvature perturbations
and it was investigated by using WMAP 3 year data~\cite{Hikage:2008sk} 
and more recently WMAP 7 year data~\cite{Hikage:2012be}.
The resultant $2\sigma$ constraint is 
\begin{equation}
   \left |f_{\rm NL}^{S,SS}\right | =  |\alpha^2 f_S| < 140,
\end{equation}
from which we obtain the following constraint:
\begin{equation}
   \left[ \theta_a^2 +\left(\frac{H_{\rm inf}}{2\pi F_a}\right)^2\right]
   \left(\frac{H_{\rm inf}}{2\pi F_a}\right)^4
   \left(\frac{F_a}{10^{12}{\rm GeV}}\right)^{3.56} < 2.5\times 10^{-17}.
\end{equation}
In Figure~\ref{figure8} the cosmological constraints on 
axion models are shown in the $H_{\rm inf}-\theta_a$ plane. 

%%%%%%%%%%%%%%%%%%%%%%%%%%%%% Figure %%%%%%%%%%%%%%%%%%%%%%%%%%%%%%%%%%%%%
%%%%%%%%%%%%%%%%%%%%%%%%%%%%%%%%%%%%%%%%%%%%%%%%%%%%%%%%%%%%%%%%%%%%%%%%%%%

%%%%%%%%%%%%%%%%%%%%%%%%%%%%%%%%%%%%%
\section{Supersymmetry and axion}
\label{sec:SUSY}
%%%%%%%%%%%%%%%%%%%%%%%%%%%%%%%%%%%%%

In this section we give a brief review on supersymmetric extension of the axion model.
Since SUSY naturally solves the gauge hierarchy problem, we have a good motivation to consider
the SUSY axion model.
In extending the axion model into a SUSY framework, there are some non-trivial features
which must be addressed carefully.
First, in SUSY, there exists a flat direction in the PQ scalar potential in addition to the massless axion.
This is due to the holomorphic property of the superpotential~\cite{Kugo:1983ma} : 
in the U(1)$_{\rm PQ}$ transformation $\phi_j \to e^{i\alpha_j} \phi_j$, the rotation parameter
$\alpha_j$ can be complex.
Therefore, by taking $\alpha_j$ to be pure imaginary, the theory should be invariant 
under the scale transformation, which means that there is a flat direction in the scalar potential correspondingly
and hence the PQ scalar is not stabilized.
This argument breaks down under the SUSY breaking effect.
Thus in order to stabilize the PQ scalar at an appropriate scale, we are forced to consider SUSY breaking 
and its effects on the structure of the scalar potential.
Second, there appear very weakly coupled massive particles, 
saxion (scalar partner of axion) and axino (fermion partner of axion), which might have
significant cosmological effects~\cite{Tamvakis:1982mw,Rajagopal:1990yx}.

%%%%%%%%%%%%%%%%%%%%%%%%%%%%%%%%%%%%%
\subsection{Stabilization of the PQ scalar}   \label{sec:stab}
%%%%%%%%%%%%%%%%%%%%%%%%%%%%%%%%%%%%%

In this subsection we show some explicit models of PQ scalar stabilization.
We do not aim to make a complete list of stabilization mechanism, but rather show typical examples,
which may be a good starting point for considering phenomenology of the SUSY axion model.

%%%%%%%%%%%%%%%%%%%%%%%%%%%%%%%%%%%%%
\subsubsection{Model A}
%%%%%%%%%%%%%%%%%%%%%%%%%%%%%%%%%%%%%

One of the simplest model for stabilizing the PQ scalar is described by the following superpotential :
\begin{equation}
	W_{\rm PQ} = \kappa X(\phi \bar\phi - \eta^2),  \label{WPQ}
\end{equation}
where $X$ is a singlet superfield with no PQ charge, and $\phi$ $(\bar\phi)$ is a PQ scalar
with PQ charge $+1$ $(-1)$, and $\kappa$ is a constant taken to be real hereafter.
The $R$-symmetry under which $X$ has a charge $+2$ and $\phi$, $\bar\phi$ are singlet
ensures this form of the superpotential.
By taking into account the SUSY breaking effect, the (relevant portion of) scalar potential is given by
\begin{equation}
	V = m_{s}^2|\phi|^2 + m_{\bar s}^2|\bar\phi|^2
	+ \kappa^2 |\phi\bar\phi-\eta^2|^2 + \kappa^2 |X|^2\left( |\phi|^2+|\bar\phi|^2 \right).
\end{equation}
where $m_{s}^2$ and $m_{\bar s}^2$ represent soft SUSY breaking masses for $\phi$ and $\bar\phi$.
PQ scalars are stabilized at $\langle \phi\rangle \sim \langle \bar\phi\rangle \sim \eta$ and $\langle X\rangle\sim 0$.
To be more precise, $X$ obtains a VEV of $\langle X\rangle\sim m_{3/2}/\kappa$,
where $m_{3/2}$ denotes the gravitino mass,
by taking account of the constant term in the superpotential $(W_0 = m_{3/2}M_P^2)$ to cancel the
cosmological constant.
In this model, the saxion corresponds to the fluctuation around the VEVs of $\phi$ and $\bar\phi$
along the flat direction $\phi\bar\phi=\eta^2$, and it obtains a mass of soft SUSY breaking scale,
which might be of the order of TeV.
The axino ($\tilde a$) corresponds to the light combination of $\tilde\phi$ and $\tilde{\bar\phi}$,
which are fermionic components of $\phi$ and $\bar\phi$, and obtains a mass of $m_{\tilde a}=\kappa \langle X \rangle \simeq m_{3/2}$.
(The other combination mixes with $\tilde X$ and gets a mass of $\kappa \eta$.)\footnote{
	Recently it was proposed that the PQ scalars in superpotential~\ref{WPQ} can take a role of waterfall field in 
	hybrid inflation, while $X$ is the inflaton~\cite{Kawasaki:2010gv,Kawasaki:2011ym,Kawasaki:2012wj}.
	A right amount of density perturbation is reproduced for $F_a \sim 10^{15}$\,GeV.
}

In order to solve the strong CP problem, we introduce a superpotential 
\begin{equation}
	W_{\rm KSVZ} = k \phi Q\bar Q, \label{WKSVZ}
\end{equation}
for the SUSY version of KSVZ axion model, where $Q$ and $\bar Q$ transforms as
fundamental and anti-fundamental representations of SU(3)$_c$,\footnote{
	To maintain the gauge coupling unification, 
	it is often assumed that $Q$ and $\bar Q$ are complete multiplets of SU(5) GUT gauge group.
}
and both have PQ charges of $-1/2$.
On the other hand, in the SUSY version of DFSZ model, we introduce
\begin{equation}
	W_{\rm DFSZ} = \lambda \frac{\phi^2}{M_P} H_u H_d, \label{WDFSZ}
\end{equation}
where $H_u$ and $H_d$ denote the up- and down-type Higgs doublets. 
In this case, $(H_u H_d)$ has a PQ charge of $-2$ and hence (some of) MSSM fields also have PQ charges.
After $\phi$ obtains a VEV, equation~\ref{WDFSZ} gives the higgsino mass, so-called $\mu$-term, as
$\mu = \lambda \eta^2 / M_P$ and it is of the order of TeV for $\lambda = O(1)$ and $F_a \sim 10^{11}$\,GeV.
Thus the PQ scale may be related with the solution to the $\mu$-problem in MSSM~\cite{Kim:1983dt}.

%%%%%%%%%%%%%%%%%%%%%%%%%%%%%%%%%%%%%
\subsubsection{Model B}
%%%%%%%%%%%%%%%%%%%%%%%%%%%%%%%%%%%%%

Let us consider the following superpotential,
\begin{equation}
	W_{\rm PQ} = \frac{\phi^n \bar\phi}{M^{n-2}}, \label{WNR}
\end{equation}
where $M$ denotes the cutoff scale and $n (\geq 3)$ is an integer.
The PQ scalars $\phi$ and $\bar\phi$ have PQ charges of $+1$ and $-n$, respectively.
The scalar potential after including the SUSY breaking effect is given by~\cite{Murayama:1992dj}
\begin{equation}
	V = -m_{s}^2|\phi|^2-m_{\bar s}^2|\bar\phi|^2 + \left(A_\phi \frac{\phi^n \bar\phi}{M^{n-2}} +{\rm h.c.}  \right)
	+ \frac{|\phi|^{2(n-1)}}{M^{2(n-2)}}\left(|\phi|^2+n^2|\bar\phi|^{2}\right),
\end{equation}
where $A_\phi$ denotes the $A$-term contribution to the SUSY breaking potential.
Here the soft mass squared of $\phi$ is assumed to have tachyonic form.
Then $\phi$ is stabilized at
\begin{equation}
	\eta=  \langle\phi\rangle \sim \left( m_s M^{n-2} \right)^{1/(n-1)}.
\end{equation}
For example, for $n=3$ and $M=M_P$, we have $\eta \sim \sqrt{m_sM_P}$
and it is of the order of $10^{11}$\,GeV for the soft mass $m_s \sim 1$\,TeV.
The $A$-term also induces VEV of $\bar\phi$ as $\langle\bar\phi\rangle \sim A_\phi\eta / m_s$,
which is smaller than $\eta$ if $A_\phi < m_s$.
The axino consists of mixtures of $\tilde\phi$ and $\tilde{\bar\phi}$, whose mass is given by
$m_{\tilde a} \sim m_s$, as seen by substituting VEVs of $\phi$ and $\bar\phi$ into Equation \ref{WNR}.
Note that the axino can be much heavier than the gravitino if $m_s \gg m_{3/2}$ due to gauge-mediation effect (see below).

%%%%%%%%%%%%%%%%%%%%%%%%%%%%%%%%%%%%%
\subsubsection{Model C}
%%%%%%%%%%%%%%%%%%%%%%%%%%%%%%%%%%%%%

In the above two models, we introduced two (or possibly more) PQ scalar fields for giving rise to the scalar potential. PQ scalars are stabilized by the combination of SUSY potential and SUSY breaking potential.
However, there may be multiple sources of SUSY breaking effects.
In the gauge-mediated SUSY breaking (GMSB) model, the dominant contribution to the
soft SUSY breaking mass often comes from the GMSB effect, while small but non-negligible 
gravity-mediation effect generally exists.
In particular, for a gauge-singlet scalar such as the PQ scalar, the effect of GMSB on its potential
is rather non-trivial because it arises at higher loop level~\cite{ArkaniHamed:1998kj,Asaka:1998ns,Asaka:1998xa}.

Let us consider the KSVZ model and see how the scalar potential is generated
through the SUSY breaking effect.
For simplicity, we adopt the following messenger sector 
\begin{equation}
	W_{\rm mess} = \kappa X \Psi \bar\Psi,
\end{equation}
where $X$ denote the SUSY breaking field and $\Psi$ and $\bar\Psi$ are messengers fields.
For $k|\phi| \ll M_{\rm mess}$, where $M_{\rm mess}$ is the messenger scale,
the mass splitting on the scalar and fermionic components of $Q, \bar Q$ yield correction to the saxion mass.
In the opposite case $k|\phi| \gg M_{\rm mess}$, after integrating out the PQ quarks,
the K\"ahler potential of the $X$ field below the PQ scale is given by
\begin{equation}
	\mathcal L = \int d^4\theta Z_X(|\phi|) |X|^2,
\end{equation}
where the wave-function renormalization factor $Z_X$ logarithmically depends on $|\phi|$ at the three-loop level.
Thus we can estimate the PQ scalar potential as
\begin{equation}
	|\phi|\frac{\partial V^{\rm (GM)}}{\partial |\phi|} \simeq 
	\left\{
	\begin{array}{ll}
%		\displaystyle
        -\frac{4k^2}{\pi^2}m_{s}^2|\phi|^2  & {\rm for}~~~k|\phi| \ll M_{\rm mess} \\
%		\displaystyle
        -\frac{g_s^4 \kappa^2}{(4\pi^2)^3} |F_X|^2
		\log^2\left( \frac{k|\phi|}{M_{\rm mess}}\right) & {\rm for}~~~k|\phi| 
		\gg M_{\rm mess} .
	\end{array}
	\right.
	\label{VGMSB}
\end{equation}
where $m_{s}\equiv(g_s^2/16\pi^2)|F_X/X|$ is the typical soft mass induced by the GMSB effect.
This potential has a negative slope, and hence PQ scalar is driven away from the origin.
It can be stopped by other contributions.
Since $\phi$ also feels the gravity-mediated SUSY breaking effect, there arises term
\begin{equation}
	V^{\rm (grav)} \simeq m_{3/2}^2 |\phi|^2.
	\label{Vgrav}
\end{equation}
The PQ scalar can be stabilized by the balance between Equations \ref{VGMSB} and \ref{Vgrav}~\cite{Asaka:1998ns} as
\begin{equation}
 \eta = \langle|\phi|\rangle \simeq \frac{g_s^2 \kappa}{8\pi^3}\frac{|F_X|}{m_{3/2}},
\end{equation}
if $k\eta > M_{\rm mess}$. If $X$ dominantly breaks SUSY, i.e., $|F_X|=\sqrt{3}m_{3/2}M_P$,
we need small $\kappa$ in order to obtain intermediated PQ scale.
In this model, the axino is expected to obtain mass dominantly through radiative effect, which is evaluated as
$m_{\tilde a} \simeq (k^2/16\pi^2)A_Q$, where $A_Q$ denotes the $A$-term contribution to the KSVZ coupling :
$\mathcal L = A_Q k \phi \tilde Q \tilde{\bar Q} +{\rm h.c.}$ with $\tilde Q (\tilde{\bar Q})$ denoting the scalar component.
%\footnote{
%	Non-renormalizable terms in the K\"ahler potential as $K \sim |\phi|^4(z+z^\dagger)/M_P^3$,
%	where $z$ is the SUSY breaking field (which may be identified with $X$),
%	yields the axino mass of $m_{\tilde a}\sim m_{3/2}\eta^2/M_P^2$. This is sufficiently small to be neglected.
%	The term $K \sim |\phi|^4/M_P^2$ yields $m_{\tilde a}\sim |F_\phi| \eta/M_P^2$, which is also the same order as the above.
%}

One can construct a variant of above mentioned models.
For example, one can make use of the scalar potential induced by
the term~\ref{WNR} to stabilize the PQ scalar against the potential~\ref{VGMSB}~\cite{Banks:2002sd,Choi:2011rs,Nakayama:2012zc}.
See also \cite{Abe:2001cg,Nakamura:2008ey,Jeong:2011xu} for ideas to radiatively
stabilize the PQ scalar.
%It is also possible that, by allowing the mixing between PQ quarks and messengers,
%the scalar potential for the PQ scalar is generated at the one-loop level~\cite{Jeong:2011xu}.

%%%%%%%%%%%%%%%%%%%%%%%%%%%%%%%%%%%%%
\subsection{Axino cosmology}
%%%%%%%%%%%%%%%%%%%%%%%%%%%%%%%%%%%%%

The axinos are produced in the early Universe and they might cause cosmological disasters
since the axino has a long lifetime.
In the KSVZ model, the dominant axino production process is scattering between gluons and gluinos
through the axino-gluino-gluon interaction.
The axino abundance, in terms of its number-to-entropy ratio, was evaluated as~\cite{Covi:2001nw,Brandenburg:2004du,Strumia:2010aa}
\begin{equation}
	Y_{\tilde a}^{(g)} \equiv \frac{n_{\tilde a}^{(g)}}{s} \simeq 2\times 10^{-6} g_s^6
	%\ln \left( \frac{1.108}{g_s} \right)
	\left( \frac{F_a}{10^{11}\,{\rm GeV}} \right)^{-2}
	\left( \frac{T_{\rm R}}{10^{5}\,{\rm GeV}} \right).
	\label{Yag}
\end{equation}
where $T_{\rm R}$ denotes the reheating temperature after inflation.
In the DFSZ model, there is an axino-Higgs-higgsino interaction.
Since this is Yukawa coupling with a coupling constant of $\sim \mu/F_a$, the axino production rate
through the Higgs and higgsino scatterings increases at low temperature ($T\sim \mu$).
The axino abundance in the DFSZ model in such a process is evaluated recently as~\cite{Chun:2011zd,Bae:2011jb,Bae:2011iw}
\begin{equation}
	Y_{\tilde a}^{(h)}\simeq 10^{-5}
	%\ln \left( \frac{1.108}{g_s} \right)
	\left( \frac{F_a}{10^{11}\,{\rm GeV}} \right)^{-2}
	\left( \frac{\mu}{1\,{\rm TeV}} \right)^2.
	\label{Yah}
\end{equation}
This contribution does not depend on the reheating temperature $T_{\rm R}$.

Cosmological implication of the axino depends on whether it is the lightest SUSY particle (LSP) or not.\footnote{
	Evaluation of the axino mass is rather a delicate issue~\cite{Goto:1991gq,Chun:1992zk,Chun:1995hc}.
	See also \cite{Higaki:2011bz,Kim:2012bb} for recent calculations.
}
If it is not the LSP, the axino decays into the neutralino whose abundance is limited from the observed DM abundance.
If the neutralino annihilation cross section is fairly large, as in the case of higgsino or wino like neutralino,
the axino decay can produce a right amount of neutralino for appropriate decay temperature of the 
axino~\cite{Choi:2008zq,Baer:2011uz}.

If the axino is the LSP, on the other hand, it can also be produced by the late decay of heavier SUSY particles.
Actually, as shown previously, the axino mass is comparable to the gravitino mass, or it can be much lighter.
In this case the axino abundance depends on the sparticle abundance after their freeze out.
Thus if the sparticle annihilation cross section is not so small (i.e., if it would reproduce the DM abundance
if it were the LSP), no significant constraint is imposed.
Axinos are also produced by the gravitinos, which decay into axinos and axions~\cite{Asaka:2000ew}.
In this case the resulting axino abundance depends on the gravitino abundance,
which may be produced thermally or non-thermally (e.g., by the decay of inflaton).
If the axino is much lighter than the gravitino, the constraint on the reheating temperature can be relaxed.

%%%%%%%%%%%%%%%%%%%%%%%%%%%%%%%%%%%%%
\subsection{Saxion cosmology}
%%%%%%%%%%%%%%%%%%%%%%%%%%%%%%%%%%%%%

Next let us consider the cosmological effects of the saxion.
Saxions are produced thermally in a similar way to the axino, but 
there also exists a contribution from the coherent oscillation.
The evaluation of the latter contribution is quite model dependent.
As shown in previous section, there are many models of saxion stabilization,
and the scalar field dynamics should be calculated in each PQ stabilization model.
We here give a typical behavior of the saxion dynamics, but one should keep in mind that
the dynamics can be far more complicated in some concrete models,
which would result in orders of magnitude difference in the estimate of saxion abundance.

First, in the KSVZ model,
the abundance of thermally produced saxions through the scattering by the gluons is given by~\cite{Graf:2012hb}
\begin{equation}
	Y_{\sigma}^{(g)} \equiv \frac{n_{\sigma}^{(g)}}{s} \simeq 2\times 10^{-6} g_s^6
	%\ln \left( \frac{1.108}{g_s} \right)
	\left( \frac{F_a}{10^{11}\,{\rm GeV}} \right)^{-2}
	\left( \frac{T_{\rm R}}{10^{5}\,{\rm GeV}} \right).
	\label{Ys}
\end{equation}
As far as we have recognized, no detailed calculations were performed 
for the saxion abundance created by the saxion-higgsino or saxion-Higgs interactions in the DFSZ model
(denoted by $Y_\sigma^{(h)}$ hereafter),
but we reasonably expect that it is same order as the corresponding axino abundance 
(Equation~\ref{Yah}).

As for the coherent oscillation, we first focus on the Model A in Sec.~\ref{sec:stab}.
In this case, the saxion feels the Hubble-induced mass in the early Universe
and hence the saxion sits at the minimum determined by the Hubble mass if the Hubble-induced mass squared has positive coefficients.
At $H\sim m_s$, the saxion begins to oscillate around the true minimum with initial amplitude $\sigma_i$.
The abundance is thus given by
\begin{equation}
	\frac{\rho_\sigma^{\rm (CO)}}{s} = 
	\left\{
	\begin{array}{ll}
%	\displaystyle
		\frac{1}{8}T_{\rm R}\left( \frac{\sigma_i}{M_P} \right)^2 
		\simeq 2\times 10^{-11}\,{\rm GeV}\left( \frac{T_{\rm R}}{10^5\,{\rm GeV}} \right)
		\left( \frac{F_a}{10^{11}\,{\rm GeV}} \right)^2
		\left( \frac{\sigma_i}{F_a} \right)^2
		&{\rm for~~}  T_{\rm R} \lesssim T_{\rm os} \\
%	\displaystyle
		\frac{1}{8}T_{\rm os}\left( \frac{\sigma_i}{M_P} \right)^2  
		\simeq 2\times 10^{-7}\,{\rm GeV}\left( \frac{m_\sigma}{1\,{\rm GeV}} \right)^{1/2}
		\left( \frac{F_a}{10^{11}\,{\rm GeV}} \right)^2
		\left( \frac{\sigma_i}{F_a} \right)^2
		&{\rm for~~}  T_{\rm R} \gtrsim T_{\rm os}
	\end{array}
	\right.
\end{equation}
where $T_{\rm os} \equiv (10/\pi^2 g_*)^{1/4}\sqrt{m_\sigma M_P}$.
Note that it does not take into account various effects which would modify the abundance significantly.
For example, thermal effects modify the $\phi$ potential in the KSVZ model,
which can change the saxion dynamics significantly~\cite{Kawasaki:2010gv,Kawasaki:2011ym,Moroi:2012vu}.
Thermal corrections to the scalar potential are negligible in the DFSZ model.
On the other hand, if the Hubble-mass squared has negative coefficient, the saxion is driven away to $\sim M_P$
during inflation, hence we expect $\sigma_i \sim M_P$ in this case.\footnote{
	In such a case $(\sigma_i\sim M_P)$, the axion isocurvature perturbation is suppressed,
	because it is proportional to $\delta\theta/\theta = \delta\theta_i/\theta_i \sim H_{\rm inf}/\sigma_i$~\cite{Kawasaki:2008jc}.
	On the other hand, if the coefficients are mildly positive, it is possible to create the axion isocurvature perturbation with
	extremely blue spectrum by considering the dynamics of saxion during inflation~\cite{Kasuya:2009up}.
}

\begin{figure}
  \begin{center}
  \hspace{0mm}\scalebox{.7}{\includegraphics{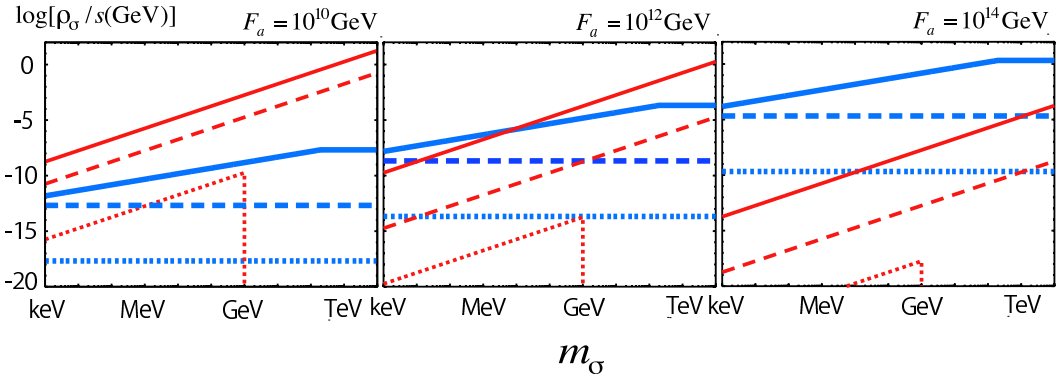}} 
  \end{center}
  \caption{
  	The saxion abundance as a function of saxion mass $m_\sigma$ for $F_a=10^{10}, 10^{12}, 10^{14}$\,GeV from left to right.
	Thick blue lines correspond to $(\rho_\sigma /s)^{\rm (CO)}$  with $\sigma_i=F_a$
	and thin red lines to $m_\sigma Y_\sigma^{(g)}$ in the KSVZ model.
	Solid, dashed and dotted lines correspond to $T_{\rm R}=10^{10},10^5$ and $1$\,GeV respectively. 
  }
  \label{figure9}
\end{figure}

The saxion abundance is given by sum of thermal contribution and coherent oscillation :
$\rho_\sigma /s = (\rho_\sigma /s)^{\rm (CO)}+m_\sigma\left( Y_\sigma^{(g)}  +Y_\sigma^{(h)} \right) $.
Figure~\ref{figure9} shows it for the KSVZ model.
The saxion abundance is bounded above from cosmological/astrophysical arguments 
depending on the saxion lifetime.
Here we list relevant constraints depending on the saxion lifetime 
$\tau_\sigma$~\cite{Kawasaki:2007mk}.

%%%%%%%%%%%%%%%%%%%%%%%%%%%%% Figure %%%%%%%%%%%%%%%%%%%%%%%%%%%%%%%%%%%%%
%%%%%%%%%%%%%%%%%%%%%%%%%%%%%%%%%%%%%%%%%%%%%%%%%%%%%%%%%%%%%%%%%%%%%%%%%%%

\begin{figure}
  \begin{center}
  \hspace{0mm}\scalebox{.7}{\includegraphics{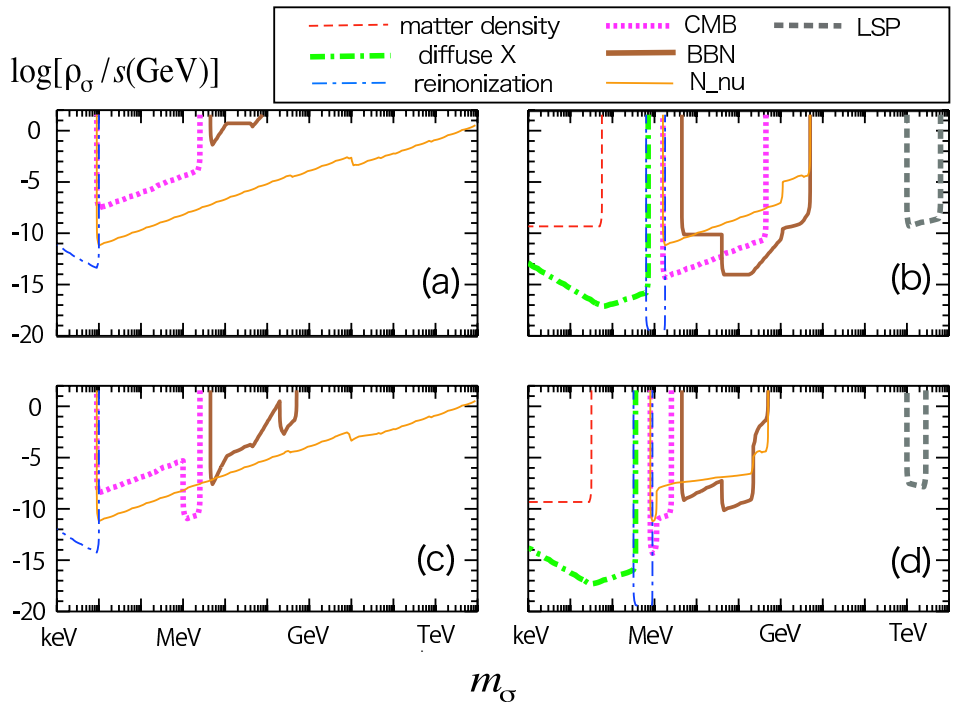}} 
 \end{center}
  \caption{
  	Cosmological constraints on the saxion abundance as a function of saxion mass $m_\sigma$ for $F_a=10^{10}$\,GeV
	in the KSVZ model ((a) and (b)) and DFSZ model ((c) and (d)).
	In (a) and (c), the saxion decay into axions is assumed to be unsuppressed.
	In (b) and (d), it is assumed to be suppressed.
	See \cite{Kawasaki:2007mk} for more detail.
  }
  \label{figure10}
\end{figure}

\begin{itemize}
\item $\tau_\sigma\lesssim 10^{12}\,{\rm sec}$ : The axion produced by the saxion decay
should not contribute to the effective number of neutrino species too much.
Conversely, it can explain the recent observational claim of the existence of dark radiation~\cite{Ichikawa:2007jv,Kawasaki:2011ym,Kawasaki:2011rc,Jeong:2012np,Choi:2012zna}.

\item $1\,{\rm sec}\lesssim \tau_\sigma \lesssim 10^{12}\,{\rm sec}$ : Injected visible energy by the saxion decay
should not alter the abundance of light elements synthesized by the big-bang nucleosynthesis (BBN).

\item $10^6\,{\rm sec}\lesssim \tau_\sigma \lesssim 10^{12}\,{\rm sec}$ : Injected visible energy by the saxion decay
should not distort the blackbody spectrum of CMB.

\item $\tau_\sigma \gtrsim 10^{12}\,{\rm sec} $ : Photons produced by the saxion decay should not exceed the observed
diffuse extragalactic background photon spectrum.

\item $\tau_\sigma \gtrsim 10^{12}\,{\rm sec} $ : Injected photons should not ionize the neutral hydrogen too much,
which would otherwise leads to too early epoch of reionization.

\item $\tau_\sigma \gtrsim 10^{17}\,{\rm sec} $ : The saxion abundance itself should not exceed the observed DM abundance.

\end{itemize}
These set bound on the saxion abundance and reheating temperature for every mass range of the saxion.
Figure~\ref{figure10} shows cosmological constraints on the saxion abundance for a wide range of the saxion mass.
For more details on these constraints, see \cite{Kawasaki:2007mk} and references therein.

%%%%%%%%%%%%%%%%%%%%%%%%%%%%% Figure %%%%%%%%%%%%%%%%%%%%%%%%%%%%%%%%%%%%%
%%%%%%%%%%%%%%%%%%%%%%%%%%%%%%%%%%%%%%%%%%%%%%%%%%%%%%%%%%%%%%%%%%%%%%%%%%%

For Models B and C, the saxion is likely trapped at the origin $\phi=0$ due to thermal effects in KSVZ-type models.
In this case, thermal inflation~\cite{Lyth:1995ka} takes place
because of the potential energy $V(|\phi|=0)$.
After thermal inflation, the saxion oscillation dominates the Universe and its decay causes the reheating.
Cosmological implications of such a scenario were discussed in \cite{Stewart:1996ai,Choi:1996vz,Chun:2000jr,Kim:2008yu,Choi:2009qd,Park:2010qd}.
It should be noticed that the domain wall number $N_{\rm DW}$ must be equal to one in order to avoid the
axionic domain wall problem (see Section~\ref{sec:PQ_breaking_after}).

%%%%%%%%%%%%%%%%%%%%%%%%%%%%%%%%%%%%%
\section{Discussion}
\label{sec:conc}
%%%%%%%%%%%%%%%%%%%%%%%%%%%%%%%%%%%%%

We have reviewed recent developments in the field of axion cosmology and its SUSY version.
In particular, according to recent simulations on axion emissions from axionic strings and domain walls,
the upper bound on the PQ scale reads $F_a \lesssim (2.0 - 3.8)\times 10^{10}$\,GeV for $N_{\rm DW}=1$ if the PQ symmetry is broken after inflation. 
There is no room for the case of $N_{\rm DW}\geq 2$ unless the phase of explicit PQ breaking bias term is finely tuned. 
This gives a tight bound on the PQ model.
If the PQ symmetry is broken before/during inflation, this constraint does not apply.
Instead, the constraint from axion isocurvature perturbations requires low inflation energy scale.
Recent investigations on non-Gaussianity in the isocurvature perturbation has shown that it gives comparable constraint
to that from the analysis of CMB power spectrum.
We have also reviewed SUSY axion models with some explicit examples of saxion stabilization.
Recent studies on cosmology of saxion and axino are briefly discussed.

Below we list some recent topics related to the axion physics which were not covered in the main text.

\paragraph{Axion thermalization}
In a recent series of works~\cite{Sikivie:2009qn,Erken:2011vv,Erken:2011dz}
it was pointed out that the cold axion thermalizes by the gravitational interaction and forms Bose-Einstein condensate,
which also leads to drastic effects on the effective number of neutrinos through the effective photon-axion interaction.
Currently there are little discussions on this claim and more investigations will be necessary (see also \cite{Saikawa:2012uk}).

\paragraph{SUSY axion model with 125\,GeV Higgs}
After the discovery of the 125\,GeV Higgs-like scalar boson~\cite{:2012gk,:2012gu}, many models were proposed
in SUSY framework. Some of them are closely related to the axion solution to the strong CP problem.
In \cite{Nakayama:2012zc}, vector-like matter was introduced for raising the lightest Higgs mass
while the PQ symmetry plays an essential role to explain the appropriate mass of the vector-like matter and to solve the $\mu$-problem.
In \cite{Jeong:2012ma,Bae:2012am}, a singlet extension of the MSSM was considered for raising the lightest Higgs mass,
where an appropriate value for the tadpole term in the singlet sector is explained by the PQ symmetry.
Further LHC data may be able to confirm such a scenario.

\paragraph{String axion and axiverse}
Finally, we want to make a brief comment on realization of the QCD axion in the string theory.
The origin of global PQ symmetry is somewhat mysterious since even the Planck-suppressed PQ violating operators 
must be highly controlled.
In string theory, such a shift symmetry of the axion-like field often appears after the compactification of extra dimensions
in the zero-modes of the dilaton, NS-NS 2-form and 
R-R $p$-forms with $p=1,3$ in type IIA and $p=0,2,4$ in type IIB theories (see e.g.~\cite{Svrcek:2006yi}).
These fields acquire masses from non-perturbative effects which break the shift symmetry.
Since these effects exponentially depend on various parameters, axions with wide mass ranges in logarithmic scale
are expected to exist, one of which may be the QCD axion: the so-called string axiverse scenario~\cite{Arvanitaki:2009fg}.
Here one should ensure that the mechanism of saxion/moduli stabilization does not give rise to the axion mass.
This severely restricts the stabilization mechanism, and discussions on this topic are found in \cite{Conlon:2006tq,Choi:2006za,Dine:2010cr,Higaki:2011me,Cicoli:2012sz} mostly for the type IIB theory.
Once a successful saxion stabilization mechanism is identified, its cosmological effects can be discussed.
It will be an interesting topic which has not been investigated in detail so far.

%%%%%%%%%%%%%%%%%%%%%%%%%%%%%%%%%%%%%
\section*{Acknowledgment}
%%%%%%%%%%%%%%%%%%%%%%%%%%%%%%%%%%%%%

This work is supported by Grant-in-Aid for Scientific research from
the Ministry of Education, Science, Sports, and Culture (MEXT), Japan,
No.\ 14102004 (M.K.), No.\ 21111006 (M.K. and K.N.), No.\ 22244030 (K.N.) and also 
by World Premier International Research Center
Initiative (WPI Initiative), MEXT, Japan.

%%% Numbered Literature Cited

%% Caution: Not all Annual Reviews series use this format for
%% bibliography entries.Your Production Editor will advise you
%% on correct format for your particular series.

\end{document}